\documentclass[]{aa} 
\usepackage{latexsym}
\usepackage{amssymb}
\usepackage{graphicx}
\usepackage{psfig}
\usepackage{natbib}
\usepackage{txfonts}
\def\etal{et al.}

\def\av{A$_{\rm v}$}

\def\teff{\ifmmode T_{\rm eff} \else $T_{\mathrm{eff}}$\fi}

\def\ltsima{$\buildrel<\over\sim$}
\def\lsim{\lower.5ex\hbox{\ltsima}}

\newcommand{\hii}{H~{\sc ii}}
\newcommand{\ha}{\ifmmode {\rm H}\alpha \else H$\alpha$\fi}
\newcommand{\hb}{\ifmmode {\rm H}\beta \else H$\beta$\fi}
\newcommand{\lya}{\ifmmode {\rm Ly}\alpha \else Ly$\alpha$\fi}

\def\micron{$\mu$m}

\def\ergscm{erg s$^{-1}$ cm$^{-2}$}
\def\msun{\ifmmode M_{\odot} \else M$_{\odot}$\fi}
\def\msunyr{\ifmmode M_{\odot} {\rm yr}^{-1} \else M$_{\odot}$ yr$^{-1}$\fi}
\def\zsun{\ifmmode Z_{\odot} \else Z$_{\odot}$\fi}
\def\lsun{\ifmmode L_{\odot} \else L$_{\odot}$\fi}

\def\mup{\ifmmode M_{\rm up} \else M$_{\rm up}$\fi}
\def\mlow{\ifmmode M_{\rm low} \else M$_{\rm low}$\fi}

\def\zacs{z$_{\rm 850LP}$}

\def\lbol{\ifmmode L_{\rm bol} \else L$_{\rm bol}$\fi}
\def\smm{SMMJ14009+0252}

\def\aap{A\&A}

\def\aj{AJ}
\def\apj{ApJ}
\def\apjl{ApJL}
\def\apjs{ApJS}
\def\mnras{MNRAS}
\def\pasp{PASP}
\newcommand{\oh}{\ifmmode 12 + \log({\rm O/H}) \else$12 + \log({\rm
O/H})$\fi}
\def\hyperz{{\em Hyperz}}
\def\flyf{\ifmmode f_{\rm Lyf} \else $f_{\rm Lyf}$\fi}
\def\pz{\ifmmode P(z) \else $P(z)$\fi}
\def\ki2{\ifmmode \chi^2 \else $\chi^2$\fi}
\def\zphot{\ifmmode z_{\rm phot} \else $z_{\rm phot}$\fi}
\def\zfit{\ifmmode z_{\rm fit} \else $z_{\rm fit}$\fi}

\newcommand{\vexp}{\ifmmode v_{\rm exp} \else $v_{\rm exp}$\fi}
\newcommand{\nh}{\ifmmode N_{\rm H} \else $N_{\rm H}$\fi}

\begin{document}
%
\title{EROs found behind lensing clusters. II. 
Stellar populations and dust properties of optical dropout EROs and comparison with related objects
\thanks{Based on observations collected at the Very Large Telescope (Antu/UT1), 
European Southern Observatory, Paranal, Chile 
(ESO Programs 69.A-0508, 70.A-0355, 73.A-0471), 
the NASA/ESA \textit{Hubble Space Telescope} obtained at the Space Telescope 
Science Institute which is operated by AURA under NASA contract NAS5-26555,
the Spitzer Space Telescope, which is operated by the Jet Propulsion Laboratory, 
California Institute of Technology under NASA contract 1407,
and the Chandra satellite.}}

\authorrunning{D. Schaerer et al.}
\titlerunning{Properties of optically faint or non-detected lensed EROs and related objects}

   \author{D.Schaerer
          \inst{1,2}
          \and
          A.Hempel\inst{1} \and E.Egami\inst{3}\and R.Pell\'{o}\inst{2} \and J.Richard\inst{2,4}
	\and J.-F. Le Borgne\inst{2} \and J.-P. Kneib \inst{5} \and M. Wise \inst{6}
	\and F. Boone\inst{7} 
          }

   \offprints{D.Schaerer}

   \institute{Geneva Observatory, University of Genevea,
              51, chemin des Maillettes, CH-1290 Sauverny, Switzerland\\
              \email{daniel.schaerer@obs.unige.ch}\\
              \email{angela.hempel@obs.unige.ch}
         \and
             Observatoire Midi-Pyr\'{e}n\'{e}es, Laboratoire
             d`Astrophysique, UMR 5572, 14 Avenue E.Belin, F-31400
             Toulouse, France
         \and
          Steward Observatory, University of Arizona, 933 North Cherry Street, Tucson, AZ 85721,USA
         \and
          Caltech Astronomy, MC105-24, Pasadena, CA 91125, USA
             \email{}
	\and OAMP, Laboratoire d'Astrophysique de Marseille, UMR 6110 traverse du Siphon, 
	F-13012 Marseille, France
	\and  Astronomical Institute Anton Pannekoek, Kruislaan 403, NL-1098 SJ Amsterdam, 
		The Netherlands
	\and  Observatoire de Paris, LERMA, 61 Av. de l'Observatoire, F-75014 Paris, France
             }

\offprints{daniel.schaerer@obs.unige.ch}
\date{Received 19 january 2007 / Accepted march 2007}

  \abstract
   {On the nature, redshift, stellar populations and dust properties of optically faint or 
	non-detected extremely red objects.}
   {We determine the nature, redshift, stellar populations 
	and dust properties of optically 
    faint or non-detected extremely red objects (ERO) found from our survey of the 
    lensing clusters A1835 and AC114 (Richard \etal\ 2006).
    Comparison with properties of related galaxies, such as IRAC selected EROs and a 
    $z \sim 6.5$ post-starburst galaxy candidate from the Hubble Ultra Deep Field.}
   {Using an updated version of \hyperz\ (Bolzonella \etal\ 2000) and a 
    large number of spectral templates we perform broad-band SED fitting. 
    The photometric observations, taken from
    Hempel \etal\ (2007), include deep optical, ACS/HST, ISAAC/VLT, IRAC/Spitzer data,
    and for some objects also 24 \micron\ MIPS/Spitzer and sub-mm data.}
   {For most of the lensed EROs we find photometric redshifts showing a strong degeneracy 
between ``low-$z$'' ($z \sim$ 1--3) and high-$z$ ($z \sim$ 6--7).
Although formally best fits are often found at high-$z$, their resulting
bright absolute magnitudes, the number density of these objects, and in some
cases Spitzer photometry or longer wavelength observations, 
suggest strongly that all of these objects are at ``low-$z$''.
The majority of these objects are best fitted with relatively young 
($\la$ 0.5--0.7 Gyr) and dusty starbursts. 
Three of our objects show indications for strong extinction,
with $A_V \sim$ 2.4--4. 
The typical stellar masses of our objects are $M_\star \sim (0.5-5) \times 10^{10} \msun$
after correction for lensing; for the most extreme ERO in our sample, the sub-mm galaxy \smm\
most likely at $\zfit \sim 3$, we estimate $M_\star \sim 6. \times 10^{11} \msun$. 
For dusty objects star formation rates (SFR) have been estimated from the bolometric luminosity
determined after fitting of semi-empirical starburst, ERO, and  ULIRG templates.
Typically we find SFR $\sim (1-18)$ \msunyr. Again, \smm\ stands out as a LIRG
with 
SFR $\sim$ 1000 \msunyr.
Finally, we predict the mid-IR to sub-mm SED of the dusty objects for 
comparison with future observations with APEX, Herschel, and ALMA.

Concerning the comparison objects, we argue that the massive post-starburst
$z \sim 6.5$ galaxy candidate HUDF-J2 showing observed properties very similar to our EROs,
is more likely a dusty starburst at $z \sim$ 2.3--2.6.
This interpretation also naturally explains the observed 24 \micron\
emission from this object and we predict its IR to sub-mm SED.

Both empirically and from our SED fits we find that the IRAC selectec EROs from 
Yan \etal\ (2004) show very similar properties to our lensed EROs.
Reasonable fits are found for most of them with relatively young and dusty
stellar populations.
}
   {}


   \maketitle
%

\section{Introduction}
Various searches for distant galaxies based on a combination of deep broad-band optical
and near-IR imaging using the Lyman break technique have been undertaken
during the last years, e.g.\ based on data in the Hubble Ultra Deep Field (HUDF),
on the GOODS survey, or others
(e.g.\ Stanway \etal\ 2003, Yan \etal\ 2003, Bouwens \etal\ 2006). 
Relying on the use of strong gravitational lensing provided by rich foreground
galaxy clusters, we have recently undertaken such a pilot program with the main
aim of identifying $z \sim$ 6--10 star forming galaxies (see e.g.\ Pell\'o \etal\ 2004, Richard
\etal\ 2006, and an overview in Schaerer \etal\ 2006).
Such candidates are selected through the now classical Lyman break technique
as optical drop-out galaxies showing an intrinsically blue UV restframe colour,
as measured from near-IR colours.

As a ``by-product'' galaxies with red spectral energy distributions (SEDs) are
also found among the drop-outs. For example, in the study of two lensing clusters
A1835 and AC114 by Richard \etal\ (2006) we found 8 lensed galaxies with $R-Ks>5.6$ 
and red near-IR 
colours satisfying one of the often used criteria of ``Extremely Red Objects'' or EROs.
The present paper focuses on these galaxies and related objects, using 
new ACS/HST observations in the \zacs\ band and Spitzer imaging obtained recently and discussed in 
Hempel \etal\ (2007, hereafter paper I).
Down to the available depth ($I_{\rm AB} \sim 27.3$ to $V_{\rm AB} \sim 28.$)
none except one of these objects is detected shortward of \zacs\ 
($\lambda_{\rm eff} \sim 9100$ \AA).
This could imply that some of them are red galaxies at very high redshift ($z \ga 6$)
or lower redshift objects with a very strong extinction.
Quantifying the properties of these EROs and comparing them with
similar objects is the main aim of the present work.
 
Generally speaking, at $z \ga 1$ EROs are found be to either dusty starbursts or old passive galaxies.
They are interesting in their own right and in the context of galaxy formation and 
evolution (e.g.\ review by McCarthy 2004).
While normally EROs are found at relatively low redshift ($z \sim$ 1--2), there
are attempts to search for similar galaxies at higher $z$ or for even more
extreme -- i.e.\ redder -- objects, by selection at longer wavelengths. 
E.g.\ Yan \etal\ (2004) have identified IRAC selected EROs (or IEROs) in the HUDF.
In these ultra-deep images the IEROs turn out to be very faint in optical 
bands ($\sim$ 27--30 mag in $V$, $i$ or \zacs), such that they could
be taken for optical drop-outs in images with less depth.
Their redshift has been estimated between $\sim$ 1.6 and 2.9 (Yan \etal).
Other selection criteria, such as searches for optical drop-out
objects with old populations 
yield at least partial overlap with 
EROs, as e.g.\ demonstrated by the post-starburst $z \sim 6.5$ galaxy
candidate of Mobasher \etal\ (2005), which also fares among the IEROs
just mentioned.
Also, a fraction of the sub-mm galaxies, detected through their strong dust emission,
show optical to near-IR colours compatible with EROs (cf.\ e.g.\ Blain \etal\ 2002).
This is also the case for one of our objects, \smm, a known lensed sub-mm galaxy
which also classifies as an ERO.
Finally, optically faint and red objects have also been found by selection of 
high X-ray to optical fluxes. These so-called EXOs are thought to be
AGN at very high redshift ($z \ga 6$) or in very dusty and/or sub-luminous 
host galaxies at more moderate redshifts ($z \sim 2-3$; see e.g.\ Koekemoer
\etal\ 2004).
 
Given this variety of optically faint or undetected and red galaxies
it is of interest to determine and compare their properties to clarify
their nature, and ultimately to obtain a coherent picture of these
seemingly different galaxy populations. 
With these objectives, we have carried out a detailed quantitative analysis 
of the stellar populations and dust properties of the lensed EROs found 
by Richard \etal\ (2006) benefiting from the new ACS/HST and Spitzer
photometry available for these lensing clusters (see paper I).
The same SED fitting method, based on the photometric redshift code \hyperz\ 
(Bolzonella \etal\ 2000) and using a large set of spectral templates, was
also applied to objects with similar SEDs, such as the IRAC selected EROs
of Yan \etal\ (2004) and other objects from the HUDF.

The paper is structured as follows.
In Sect.\ \ref{s_obs} we briefly summarise the observational data for our lensed
EROs. The SED fitting method is described in Sect.\ \ref{s_fit}.
Results for the \smm\ galaxy are presented and discussed in Sect.\ \ref{s_smm}.
The other EROs are discussed in Sect.\ \ref{s_res}.
In Sect.\  \ref{s_other} we analyse and discuss the properties of 
related objects from the HUDF.
Our main conclusions are summarised in Sect.\ \ref{s_conclude}.

Throughout this paper we adopted the following cosmology:
$\Omega_{m}=0.3$, H$_{o}$=70 km s$^{-1}$Mpc$^{-1}$ in a flat
universe. Except if mentioned otherwise, all magnitudes are 
given in the Vega system.


\begin{table*}[htb]
\caption{Photometry of the selected ERO subsample in Abell 1835 and AC 114 taken from Hempel \etal\ 
(2007, paper I). 
All magnitudes are given in the Vega system. For conversion to the AB system see the 
filter properties listed in Table \protect\ref{t_filters}.
Non-detections (lower limits) are 1 $\sigma$ values. NA stands for non available.}
\begin{tabular}{lcccccccc}
\hline 
\hline
Object     &      V   & R       & I       & \zacs\         &SZ               & J              & H              & Ks  \\
\hline
A1835-\#1  &  $>$28.1 & $>$27.8 & $>$26.7 & 25.70$\pm$0.07 & 24.44$\pm$0.27 & 22.76$\pm$0.16 & 22.40$\pm$0.08 & 20.74$\pm$0.02 \\
A1835-\#2  &  $>$28.1 & $>$27.8 & $>$26.7 & $>$27.46       & 24.08$\pm$0.26 & 24.41$\pm$1.34 & 21.78$\pm$0.06 & 20.45$\pm$0.02 \\  
A1835-\#3  &  $>$28.1 & $>$27.8 & $>$26.7 & 24.06$\pm$0.07$^a$ & 23.78$\pm$0.10 & 24.32$\pm$0.47 & 22.55$\pm$0.07 & 21.58$\pm$0.03 \\
A1835-\#4  &  $>$28.1 & $>$27.8 & $>$26.7 & 25.48$\pm$0.14 & 24.44$\pm$0.15 & 23.56$\pm$0.18 & 22.90$\pm$0.07 & 21.95$\pm$0.03 \\ 
A1835-\#10 &  $>$28.1 & $>$27.8 & $>$26.7 & 25.56$\pm$0.11 & 24.00$\pm$0.12 & 23.72$\pm$0.26 & 23.36$\pm$0.13 & 21.67$\pm$0.03 \\
A1835-\#11 &  $>$28.1 & $>$27.8 & $>$26.7 & $>$27.46       & $>$26.9        & 23.92$\pm$0.37 & 23.49$\pm$0.18 & 21.29$\pm$0.03 \\
A1835-\#17 &  $>$28.1 & $>$27.8 & $>$26.7 & $>$27.46       & $>$26.9         & $>25.6$       & 23.51$\pm$0.16 & 22.11$\pm$0.03 \\
\\
AC114-\#1  &  $>$28.5 & $>$27.7 & $>$26.8 & 24.55$\pm$0.07 & NA  & 21.26$\pm$0.04 & 19.75$\pm$0.01 & 18.62$\pm$0.001 \\
\hline
\multicolumn{9}{l}{$^a$ This source appears double in the \zacs\ image. The fainter component has 25.31$\pm$0.07.
In the SED modeling the brighter magnitude or the}\\
\multicolumn{9}{l}{sum of two have been used.}
\label{tab1}
\end{tabular}
\end{table*}

\begin{table*}[htb]
\caption{IRAC and MIPS 24 \micron\ photometry of the EROs listed in Table \ref{tab1}
taken from paper I.
All fluxes are given in $\mu$Jy.
Upper limits are 1-sigma noise values at the position of the sources.
Due to source blending the data is uncomplete, i.e.\ not available for
objects A1835-\#3, A1835-\#10, and A1835-\#11, and partially for A1835-\#1.
}
\begin{tabular}{lrllcrrrl}
\hline
\hline
       & \multicolumn{4}{c}{IRAC/Spitzer} & MIPS \\
Object & 3.6 & 4.5 & 5.8 & 8.0 & 24.0 \\
\hline
A1835-\#1  &  2.9 $\pm$ 0.2 &  4.0 $\pm$ 0.2 & \multicolumn{2}{c}{blended}     & $<$ 10 \\
A1835-\#2  & 14.4 $\pm$ 0.2 & 23.3 $\pm$ 0.3 & 37.6 $\pm$ 1.5 & 50.9 $\pm$ 1.6 & 320 $\pm$ 11 \\
A1835-\#4  &  1.9 $\pm$ 0.2 &  1.0 $\pm$ 0.3 & $<$ 0.7        & $<$ 0.6        & $<$ 10 \\
A1835-\#17 &  2.4 $\pm$ 0.2 &  1.6 $\pm$ 0.2 & $<$ 0.6        & $<$ 0.5        & $<$ 10 \\
\\
AC114-\#1  & 67.3 $\pm$ 0.5 & 64.4 $\pm$ 0.5 & 50.3 $\pm$ 1.5 & 44.9 $\pm$ 2.2 & 189.0 $\pm$ 8.9\\
\label{tobs2}
\end{tabular}
\end{table*}

\section{Observations}
\label{s_obs}
The present work deals primarily with the 8 EROs detected as red
($R-Ks > 5.6$) optically non-detected objects in the fields of the
lensing clusters A1835 and AC114 by Richard \etal\ (2006).
The images obtained by Richard \etal\ were complemented with
new ACS/HST observations in the \zacs\ band (optical), and where possible
with Spitzer observations at near- to mid-IR wavelengths (IRAC 3.6, 4.5, 5.8, and 
8.0 \micron, and MIPS 24 \micron\ photometry).

All details concerning the photometric data reduction are given in Hempel 
\etal\ (2007, paper I).
The photometry from paper I is summarised in Tables \ref{tab1}, \ref{tobs2},
and \ref{t_filters}.
In this paper the EROs from Richard \etal\ were reselected from the $Ks$
band image and optical and near-IR photometry measured with SExtractor 
using {\tt AUTO\_MAG}. 
For the photometry from different instruments source matching was done 
using object coordinates based on astrometry using the ESO-USNO-A2.0 catalog.
For Spitzer total aperture-corrected flux densities were determined from  
measurements  in 3 \arcsec\ diameter apertures.
No attempt has been made to correct the differences in flux measurement
that are caused by the different apertures and image quality. 
This is not necessary as we do not use aperture
photometry for the optical and near-infrared images, or in case of
Spitzer perform aperture correction. The photometry used, {\tt
AUTO\_MAG}, is determined by measuring the flux in a flexible
elliptical aperture around each object and accounts for the extended
brighteness distribution for the brighter objects.  
Possible differences in the photometry
with respect to the earlier measurements of Richard \etal\ (2006)
are discussed in paper I. 
Using fixed aperture near-IR photometry implies only relatively small 
changes and does not alter our overall conclusions, as test computations
have shown.
The new measurements, adapted to the morphology of the EROs, are 
used in the present paper.
In the SED modeling discussed below we consider also a minimum photometric
error to account for uncertainties due to matching photometry from different
instruments.

No object from the subsample of ``optical dropout'' EROs discussed in paper I
is detected shortward of the $R$ band. 
For modeling in the present work we include the $V$ band
non-detection as the dropout constraint. Non-detections 
at shorter wavelengths are redundant and are therefore
not included in the SED fitting.

\begin{figure}[htb]
\centering\includegraphics[width=8.8cm]{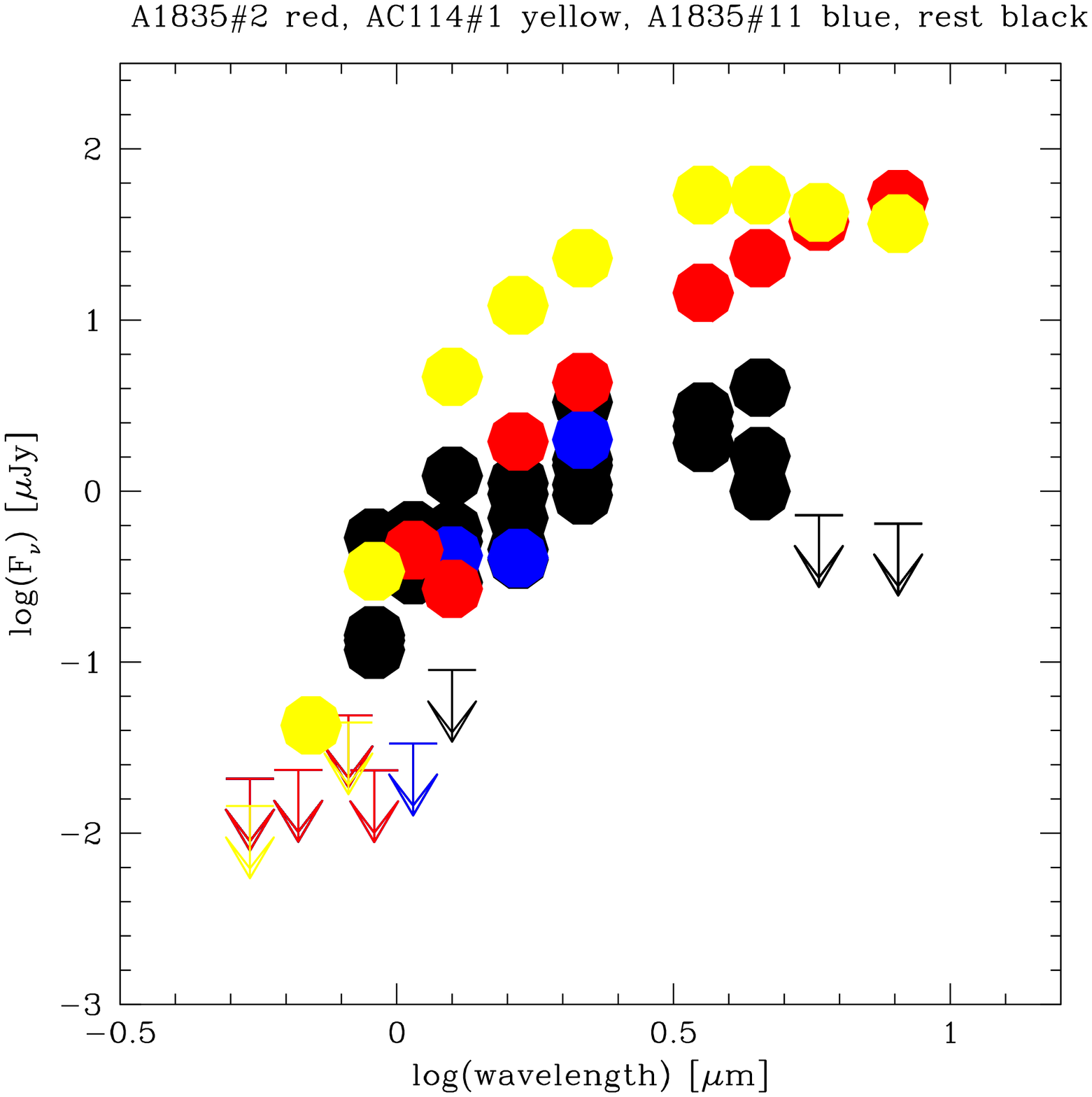}
\caption{Observed SED of all EROs in the optical and near-IR up to 10 \micron.
The two most extreme objects, the sub-mm galaxy A1835-\#2 and AC114-\#1, are indicated
in red and yellow respectively. A1835-\#11, whose properties including the photometric
redshift remain highly uncertain 
is shown in blue.}
\label{fig_allero}
\end{figure}

An overview of the observed optical, near-IR, and IRAC/Spitzer fluxes 
of all the objects discussed in this paper is shown in 
Fig.\ \ref{fig_allero}.

\section{SED fitting method}
\label{s_fit}

An SED fitting technique based on an updated version of the \hyperz\ code 
from Bolzonella \etal\ (2000) is used to
constrain the redshift, stellar population properties (age, star
formation history), and extinction of the galaxies studied in this
paper. To do so we closely follow the procedures outlined in 
Schaerer \& Pell\'o (2005). In addition we have included other
synthetic, empirical and semi-empirical spectral templates,
as described below.

\subsection{Photometry}
Ground-based, HST, and Spitzer photometry 
of the lensed EROs is taken from paper I.
The following bands have been
included in the SED fitting: $V$, $R$, $I$ nondetections, \zacs\
from ACS/HST, $SZ$, $J$, $H$, $Ks$ from ISAAC/VLT, channels 1--4
from IRAC/Spitzer and 24 \micron\ MIPS/Spitzer where available.
The filter properties are listed in Table \ref{t_filters}.

\begin{table}[htb]
\caption{Properties of the photometric filters used for the SED fitting
of objects in the field of A1835 and of AC114.
For AC114 alternate filters are listed in parenthesis and $SZ$ observations
are not available (entry NA).
Col.\ 1 indicates the filter name, col.\ 2 the 
effective wavelength in micron, col.\ 3 the effective bandpass (filter ``width'') 
in micron computed with a Gaussian approximation. 
AB corrections ($C_{AB}$), defined
by $m_{AB}=m_{\rm Vega} +  C_{AB}$, are listed in col.\ 3.
Detailed information on the photometry can be found in paper I.}
\label{t_filters}
\begin{tabular}{lcccccccc}
\hline
\hline
Filter  & $\lambda_{\rm eff}$ [$\mu$m] & $\Delta\lambda$ [$\mu$m] &$C_{AB}$ \\
\hline 
$V$               &  0.543         & 0.056  & 0.018 \\
$R$ ($R_{702}$)   &  0.664 (0.700) & 0.075 (0.123) & 0.246 (0.299) \\
$I$ ($I_{814}$)   &  0.817 (0.807) & 0.117 (0.137) & 0.462 (0.445) \\
\zacs\            &  0.911         & 0.114  & 0.540 \\
$SZ$ (NA)         &  1.070         & 0.094  & 0.698 \\
$J$               &  1.259         & 0.167  & 0.945 \\
$H$               &  1.656         & 0.180  & 1.412 \\
$Ks$              &  2.165         & 0.181  & 1.871 \\
IRAC 3.6 \micron\ &  3.577         & 0.427  & 2.790 \\
IRAC 4.5 \micron\ &  4.530         & 0.567  & 3.249 \\
IRAC 5.8 \micron\ &  5.788         & 0.801  & 3.737 \\
IRAC 8.0 \micron\ &  8.045         & 1.634  & 4.392 \\
\end{tabular}
\end{table}

For other objects, included here for comparison, the original photometry 
was taken from the literature.

In some cases a prescribed minimum photometric error is assumed, to
examine the influence of possibly underestimated error bars and to account
for uncertainties in absolute flux calibrations when combining photometry
from different instruments.

\subsection{Spectral templates}
The following spectral templates, assembled into several groups, have been used in the fit procedure.

\begin{itemize}
\item {\bf Bruzual \& Charlot} plus Coleman
\etal\ (1980) empirical templates galaxies of all Hubble types (hereafter named BC or BCCWW group).
The theoretical Bruzual \& Charlot (2001) models, taken here for solar metallicity, include various 
star formation histories representative of different Hubble types 
(burst, and e-folding times of $\tau$=1, 2, 3, 5, 15, 30, and $\infty$ Gyr corresponding
to E, S0, Sa, Sb, Sc, Sd, and Im types).
The IMF adopted in these models is the Miller-Scalo IMF from 0.1 to 125 \msun.
We have not included the templates from the Bruzual \& Charlot (2003) update, as this
concerns mostly high resolution spectral libraries which have no impact on our results.

\item {\bf Starburst SEDs from Schaerer} (2002, 2003) models at 
different metallicities extended up to ages of 1 Gyr and considering instantaneous 
bursts or constant star formation (s04gyr group).
The overall SEDs predicted by these synthesis models, including in particular nebular 
continuum emission  neglected in the Bruzual \& Charlot models, are basically 
identical to the ones from {\em Starburst99} (Leitherer \etal\ 1999). 
However, a larger variety of metallicities is included.
These models assume a Salpeter IMF from 1 to 100 \msun.

\item {\bf Maraston models} (Maraston 2005) including a semi-empirical treatment
of thermally pulsating AGB stars, whose contribution in the near-IR may be significant 
in certain phases.
This non-standard approach may therefore lead to different age and stellar mass estimates 
than other synthesis codes (see Maraston 2005 details).
The templates used here include simple stellar populations (bursts) with ages up
to 15 Gyr, and exponentially decreasing star formation histories with 
e-folding times up to 2 Gyr.

\item {\bf Empirical or semi-empirical starburst, ULIRG and QSO templates.}
In addition to starburst templates from the Calzetti et al.\ (1994) and 
Kinney et al.\ (1996) atlas  included in the public \hyperz\ version, 
we have added the HST QSO template of Zheng et al.\ (1997),  
and templates of metal-poor \hii\ galaxies SBS0335-052 and
Tol 1914-266 including numerous strong emission lines (Izotov private
communication).

To include also more obscured objects we have added 
UV to millimeter band templates of EROs, ULIRGS, starburst and normal 
galaxies (HR 10, Arp 220, M82, NGC 6090, M51, M100, NGC 6949) from fits of GRASIL models
to multi-wavelength observations (Silva \etal\ 1998, named GRASIL group). 
These templates are therefore semi-empirical templates.
This template group will be used in particular to predict mid-IR to sub-mm fluxes,
and hence to estimated total bolometric luminosities,
after fitting the optical to 8 \micron\ part of the spectrum.

\end{itemize}

\subsection{Fit procedure and determined parameters}
The main free parameters we consider are: the spectral template (among
a group), redshift $z$ , and (additional) extinction ($A_V$) assuming a
Calzetti \etal\ (2000) law. In our standard calculations $z$ is varied from 0 to
10, and \av\ from 0 to 4 mag. For test purposes higher values of \av\
and other extinction laws are also allowed.

To increase the diversity of empirical or semi-empirical templates
and to allow for possible deviations from them, reddening is optionally 
also considered as a free parameter. In this case, this obviously corresponds
to an additional reddening. 
Test computations have shown an excellent consistency between
photometric and spectroscopic redshifts using e.g. the GRASIL template
group to fit the near-IR to IRAC observations of Stern \etal\ (2006)
of the ERO HR10, once allowing for possible additional reddening.
A similar approach with empirical templates was also adopted by 
Rigby \etal\ (2005).
As for all templates the corresponding 
dust emission is not treated consistently.

Finally, from the luminosity distance of the object the scaling of the template 
SED to the observed absolute fluxes yields an absolute scaling property,
such as the stellar mass or the star formation rate (SFR) when
templates generated by evolutionary synthesis models are used.

To estimate stellar masses we use two different approaches.
{\em 1)} We determine the restframe absolute K band magnitude $M^{\rm rest}(Ks)$ from
\hyperz\ and assume a typical light-to-mass ratio $(L_K/M)$. The stellar mass is then
determined as:
\begin{equation}
	M_\star = 10^{-0.4[M^{\rm rest}(Ks)-3.3]} / (L_K/M)
\end{equation}
in solar units. The numerical value 3.3 is the solar absolute Ks band magnitude.
This approach is applicable to any template, theoretical or empirical ones.
For comparison with other dusty galaxies, we adopt the value $L_K/M=3.2$ used 
for SCUBA galaxies by Borys \etal\ (2005). 

{\em 2)} When using templates from evolutionary synthesis models, the stellar
mass (and/or SFR) can be determined from the absolute scaling of the best fit
template to the observed fluxes and to the best fit redshift.
Note that the Bruzual \& Charlot models used here assume a Miller-Scalo IMF
from 0.1 to 125 \msun, whereas the S04 models assume a Salpeter IMF from 1 to 100 \msun.

The star formation rate is a natural quantity when star formation
over a certain time scale (or at a constant rate) is considered.
This quantity can therefore only be determined through SED fits
using theoretical templates assuming constant star formation.
Alternatively, for objects with good fits using the multi-wavelength GRASIL
templates covering the restframe UV to sub-mm domain, we will determine SFRs 
from the total bolometric luminosity derived over the available spectral range
and applying a standard Kennicutt (1998) relation between $L_{\rm bol}$ and
SFR.

Absolute quantities such as the stellar mass, SFR, and bolometric
luminosity depending on the luminosity distance must also be corrected
for the effects of gravitational lensing.  The magnification factor of
each source was determined using the mass models of A1835 (similar to
Smith et al.\ 2005) and AC114 (Natarajan et al.\ 1998, Campusano et
al.\ 2001), following the same procedure as in Richard et al.\ (2006). 
Because of the slight dependence of the magnification with
the source redshift $z_s$, at the location of the EROs, we computed
different estimates assuming $z_s=0.5$, 1, 2, 3, as well as 7 for
comparison. 
The different values, given in Table  \ref{t_mu} for each object, 
reflect the uncertainty in the magnification factor, the source 
redshift being the dominant source of error.
In any case, for the bulk of these sources located somewhat away 
from the cluster center (see Figs.\ 10, 11 in Richard \etal), the 
magnification is relatively small.


\begin{table}[htb]
\caption{Magnification factors $\mu$ from the lensing models of A1835 and AC114
predicted for various source redshifts $z_s$. The values of $\mu$ are dimensionless
magnification factors, and not in magnitudes.}
\begin{tabular}{llllllllllll}
\hline
\hline
Object & $z_s=0.5$ & $z_s=1.$ & $z_s=2.$ & $z_s=3.$ & $z_s=7.$\\
\hline
A1835-\#1  & 1.13 & 1.21 & 1.26 & 1.28 & 1.33\\
A1835-\#2  & 1.38 & 1.71 & 1.95 & 2.05 & 2.32 \\
A1835-\#3  & 1.24 & 1.43 & 1.54 & 1.59 & 1.72 \\
A1835-\#4  & 1.25 & 1.45 & 1.57 & 1.62 & 1.76 \\
A1835-\#10 & 1.19 & 1.33 & 1.42 & 1.45 & 1.54 \\
A1835-\#11 & 1.14 & 1.23 & 1.28 & 1.30 & 1.35 \\
A1835-\#17 & 1.14 & 1.23 & 1.28 & 1.30 & 1.36 \\
AC114-\#1  & 1.49 & 2.24 & 2.87 & 3.15 & 4.03 \\
\end{tabular}
\label{t_mu}
\end{table}

\section{The sub-mm galaxy A1835 - \#2 (SMMJ14009+0252)}
\label{s_smm}
As already mentioned in Richard \etal\ (2006), this ERO corresponds
to the known sub-mm source SMMJ14009+0252 (Ivison \etal\ 2000,
Smail \etal\ 2002, Frayer \etal\ 2004).
With AC114-\#1 this object is the brightest optical dropout ERO from our
sample.

\subsection{SED fitting results}
\begin{figure}[htb]
\centering{\psfig{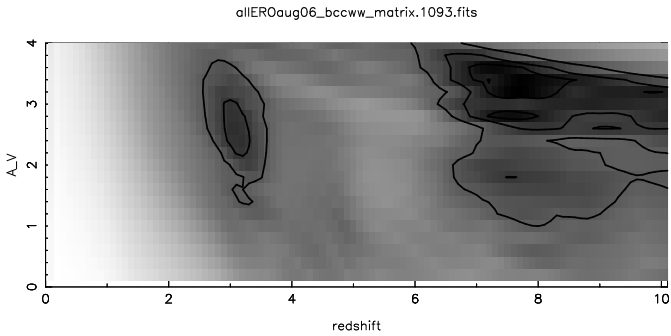}}
\caption{$\chi^2$ map 
as a function of redshift and extinction ($\chi^2(z,A_V)$) for
A1835-\#2, displaying in dark the most probable regions on a  logarithmic scale. 
Solid lines enclose the 1 to 3 $\sigma$ contours (confidence levels of 68, 90
and 99 \% respectively).
The (z,A$_V$) projection plane presented in this figure corresponds to the
best $\chi^2$ found through the model-age parameter space for templates from the BCCWW group.
Note the degeneracy of the photometric redshift solutions in this plane.
Formally the best fit is found at high redshift (\zfit $>$ 7). 
However, for various reasons, including the 24 \micron\ and sub-mm SED
and the exceptional luminosity of this object if at high $z$, the most
likely redshift of this galaxy is $\sim 3$. See discussion in text.}
\label{fig_map_2}
\end{figure}

The observed SED shows, in $F_\nu$ units (cf.\ Fig.\ \ref{fig_dust_2}), 
a continuously increasing SED from the near-IR, 
through the 4 IRAC channels and up to 24 \micron\ (MIPS), where this 
object is also detected.
When all IRAC bands are included in fits, the solutions with the
``standard'' templates are driven to high-$z$ (\zphot\ between 6 and 8).
However, a reasonably low \ki2\ is achieved for all redshifts $\zphot \ga 3$;
see \ki2\ map on Fig.\ref{fig_map_2}.
For all redshifts a very large extinction ($A_V \sim$ 3--4!) 
is found as a best fit.

\begin{figure}[htb]
\centering\includegraphics[width=8.8cm]{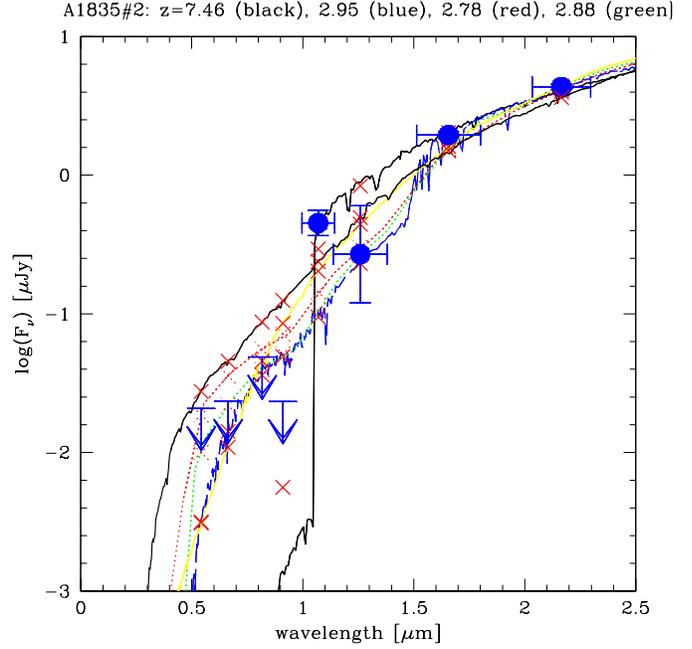}
\caption{A1835-\#2: Comparison of best fit high-z solution (black, \zfit=7.46)
with model fits at $z\sim 3$. Blue: Bruzual \& Charlot model (SF history as ellipticals
at age of 0.36 Gyr + 2.4 mag \av\ extinction). Red: GRASIL template of NGC 6090
with \av =1.4.
See discussion in text.}
\label{fig_alt_2}
\end{figure}

\begin{figure}[htb]
\centering\includegraphics[width=8.8cm]{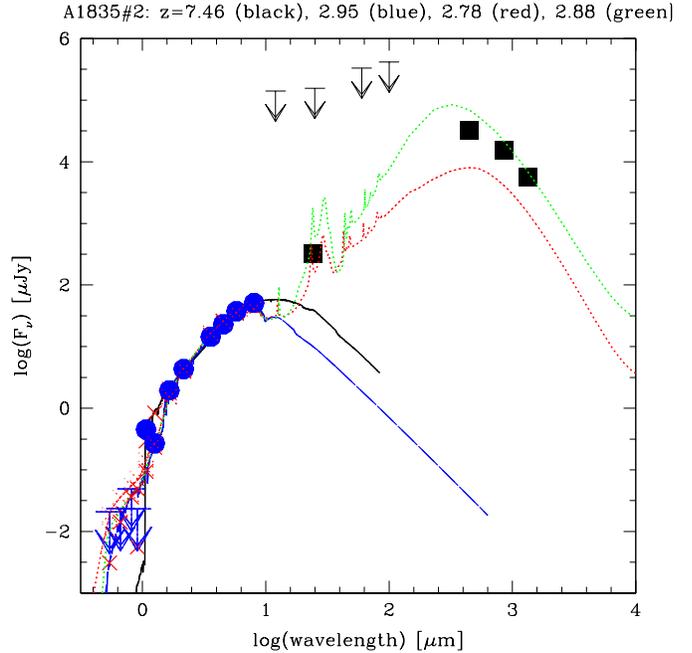}
\caption{Observed SED of the ERO/sub-mm galaxy A1835-\#2 including
VLT, Spitzer (IRAC, MIPS), SCUBA observations.
The model fits are the same as shown in Fig.\ \protect\ref{fig_alt_2}.
Note that the two fits with GRASIL models at $z \sim$ 2.8--2.9 
reproduce well the MIPS 24 \micron\ flux and bracket the observed 
sub-mm points.
}
\label{fig_dust_2}
\end{figure}

As shown in Figs.\ \ref{fig_alt_2} and \ref{fig_dust_2}, the high-$z$ solution 
(\zfit=7.46) provides an excellent fit to the observed SED.
This template corresponds to a young burst (6 Myr) + high $A_V$.
The best fit with GRASIL templates is obtained at \zfit=2.78 with
the NGC 6090 template plus additional extinction of \av=1.4.
Imposing a maximum redshift of 4 to the BCCWW templates, one finds 
a very similar best photometric redshift (\zfit=2.95) 
for an elliptical with 0.36 Gyr plus 2.4 mag extinction in \av. 
These two $z \sim 3$ solutions are also plotted in Fig.\ \ref{fig_alt_2}, 
showing a discrepancy at $\sim$ 0.95--1.1 \micron\
(cf.\ below).

Actually the overall SED of this object, including in particular our MIPS 24 \micron\
and the SCUBA measurements from Ivison \etal\ (2000), is rather well fitted with
semi-empirical templates from GRASIL for redshifts $z \sim$ 2.8--3.,
as shown on Fig.\ \ref{fig_dust_2}.
Templates with very strong dust emission such as Arp 220 are needed to reproduce the 
observed ratio of the  
sub-mm to near/mid-IR flux. For example, templates of more moderate
starbursts like M82 and NGC 6090 underpredict the sub-mm emission.
The \hyperz\ best fit with the Arp 220 template requires an additional
extinction of $A_V$ = 1.4 (for the Calzetti \etal\ law).
The only difficulty with fits at \zfit$\sim$ 3 is the excess emission observed
in the $SZ$ band, which is $\sim 5 \sigma$ above the expected level at such
redshift. A natural explanation could be the Mg~{\sc ii} $\lambda$2798 emission 
line seen in type 1 AGNs (e.g.\ Gavignaud et al.\ 2006). 

Already from the observed monotonous flux increase observed across
all 4 IRAC bands it is quite clear that this source cannot be at redshift
much smaller than $\sim$ 3. Otherwise the typical flux depression,
associated to the transition from the stellar peak at 1.6 \micron\
(restframe) to the raising dust emission at longer wavelengths
(cf.\ John 1998, Sawicki 2002) should be seen.
Excluding higher redshift solutions from the near-IR to 8 \micron\ data 
used here for the \hyperz\ SED fitting is difficult, and 
would require a more complex stellar population plus dust modeling.
However, if at $z\sim$ 6--8 as suggested from the formal best fits, the absolute
magnitude of this object would be rather exceptional 
($M^{rest}(Ks) \sim -29.6$ or $M^{rest}(V) \sim -26.6$ without correcting for lensing),
rendering this case very unlikely.
Furthermore radio and sub-mm data (cf.\ below) as well as our ``global'' 
SED analysis favour $z \ll 6$.
For these various reasons we conclude that the most likely redshift
of this object is $z \sim$ 3.
This redshift is larger than the estimate based on the radio-submm
spectral index $\alpha^{850}_{1.4}$, but in agreement with the one from
submm colours (cf.\ Ivison \etal\ 2000). It is also larger than
our previous estimate based on optical to near-IR photometry (Richard
\etal\ 2006).

Assuming $z \sim 3$ and a magnification factor $\mu=2$ (cf.\ Table \ref{t_mu})
one obtains the following estimates:
with a rest-frame absolute magnitude $M^{\rm rest}(Ks) \sim$ -27.7)
this object is $\sim$ 3.4 to 3.6 magnitudes brighter 
than $M_\star$ at this redshift (cf.\ Kashikawa \etal\ 2003) or 
than $M_\star(Ks)$ from 2MASS in the local Universe (cf.\ Kochanek \etal\ 2001).

The mass, estimated from the best fitting Bruzual \& Charlot and S04 templates
(with an age of $\sim$ 0.36 and 0.14 Gyr respectively), is 
$M_\star \sim 1.2 \times 10^{12} / \mu \msun$
\footnote{For the value $L_K/M=3.2$ adopted for SCUBA galaxies (cf.\ Borys \etal\ 2005)
one obtains $M_\star \sim 7.8 \times 10^{11} / \mu \msun$},
slightly more massive e.g.\ than the most massive SCUBA galaxy discussed by 
Borys \etal\ (2005) and typically an order of magnitude more massive
than the most massive $z \sim 3$ Lyman break galaxy observed with Spitzer
by Rigopoulou \etal\ (2006).

Integrating the global SED of Arp 220 adjusted to the observations
(see Fig.\ \ref{fig_dust_2})
and assuming $z=2.78$ one obtains a total luminosity of 
$L_{\rm bol} \sim 1.2 \times 10^{13} / \mu$ \lsun, 
close to the limit between
ultra-luminous and hyper-luminous infrared galaxies (ULIRG and HyLIRG).
Using standard SFR conversion factors (Kennicutt 1998) this corresponds to
an estimated star $SFR \sim 1050$ \msunyr,
adopting $\mu \sim 2$ (Table \ref{t_mu}).

\subsection{Discussion}
The radio-submm spectral index $\alpha^{850}_{1.4} = 0.60 \pm 0.03$
indicates a likely redshift of $0.7 \la z \la 2.3$ (Smail \etal\ 2000,
Ivison \etal\ 2000).
From the 450- to 850-\micron\ flux a coarse estimate of $z \ga 2.8$
has been derived by Hughes \etal\ (1998).
From the near-IR to submm SED and from the low value of $\alpha^{850}_{1.4}$, 
Ivison \etal\ (2000) argue that A1835-\#2 is more likely at $ 3 \la z \la 5$
and that the radio flux contains some AGN contribution.
A recent reanalysis of the radio and submm flux by Aretxaga \etal\ (2003)
yields a higher redshift estimate of $z \sim 4.1 \pm 0.8$
However, their best fit SED strongly overpredicts the observed 
near- to mid-IR flux. 
Our earlier photometric redshift estimate of $z \sim$ 1.2--1.6 (Richard
\etal\ 2006) is now superseded by the present analysis including 
in particular the longer wavelength IRAC/Spitzer observations leading
to a higher $z$.
The various redshift estimates, including the one presented here, can
be reconciled if the radio flux contains a contribution from an AGN,
as already pointed out by Ivison \etal\ (2000). 
For our best redshift, $z \sim 3$, the AGN contribution does not need
to be strong, as also discussed by these authors.
As already noted above, the observed $SZ$ band excess could also be an
indication for an AGN.

At the high luminosities of this object, in the ULIRG range, the AGN fraction
is high ($\ga$ 40--50 \%, cf.\ Veilleux \etal\ 1999, Alexander \etal\ 2004) 
rendering the AGN hypothesis quite likely.
However, is there other direct evidence for an AGN ?
Ivison \etal\ (2000) have obtained an upper limit 
in soft X-rays (0.1-2.0 keV) from ROSAT archival HRI observations.
In our recent Chandra observations of Abell 1835, described in paper I, 
this object remains undetected with flux limits of the order of
$< (2.-3.) \times 10^{-16}$ \ergscm\ in the 0.5-7.0 keV band, for 
photon power law index $\Gamma$ between 1.0 and 2.0.
The corresponding limit for 2.0-10.0 keV and $\Gamma=1.4$ is 
$\la 2.5 \times 10^{-16}$ \ergscm.
A comparison with the X-ray and 24 \micron\ fluxes of starbursts and AGN
compiled by Alonso-Herrero \etal\ (2004) places this X-ray limit
well below the typical range of hard X-ray selected AGN. A more detailed 
analysis will be needed to examine how much room these new constraints 
leave for a putative AGN in this object.

Compared to other SCUBA galaxies studied also in the rest-frame optical
(cf. Smail \etal\ 2004)
A1835-\#2 fares among the faintest ones in $K$ and among the ``reddest ones''
in optical/IR flux. 
With $m(Ks) \sim 20.5$ it is close to the faintest objects of Smail \etal,
which have $18 \la Ks \la 20.9$; 
among the 7 confirmed sub-mm galaxies observed observed in the SCUBA Cluster Lens Survey
of Frayer \etal\ (2004) it is the second faintest object in $K$ surpassed only by 
SMMJ00266+1708 with $m(K) = 22.36 \pm 0.16$, and the second reddest in $J-K$.
After lensing correction the magnitude of A1835-\#2 is $m(Ks) \sim 21.2$.
Several other sub-mm galaxies are known with very faint flux levels
at $K \ga$ 21.0--21.9 (Smail \etal\ 2002, Dannerbauer \etal\ 2002).
Its restframe V-band to IR luminosity ratio is very low, $\sim 5. \times 10^{-4}$,
placing it among the 5 most extreme sub-mm galaxies when compared to
the Smail \etal\ (2004) sample.

In terms of stellar populations we find a dominant stellar age of $\sim$ 0.36 Gyr 
or younger for A1835-\#2, similar to the mean ages of $\sim (310-530) \pm (80-90)$ Myr 
estimated by Smail \etal\
(2004) for a sample of sub-mm galaxies and optically faint radio galaxies.  
The extinction we estimate ($A_V \sim$ 2.4--3) is somewhat larger than
the average of $A_V \sim (1.70-2.44) \pm (0.13-0.14)$ found by 
Smail \etal, but comparable to the median $A_V \sim 2.9 \pm 0.5$ determined
by Takata \etal\ (2006) for sub-mm galaxies from the Balmer decrement.
The best fit with the Arp 220 spectrum, requiring an additional extinction
of $A_V =1.4$, also indicates that we're dealing with an object with
a rather exceptionally large extinction!

We note also that the stellar mass estimated from the SED fit 
($M_\star \sim 1.2 \times 10^{12} / \mu \msun$) is consistent with the mass being 
built up at the high SFR of $\sim 2100 / \mu$ \msunyr\ over a period of $\la$ 360 Myr.
This leaves room for $\sim$ 25 \% of the stellar mass being
formed from a previous star formation event.

\section{SED fitting results for other objects}
\label{s_res}

We now present the results from the SED fits for the individual objects.
First we discuss in detail the objects for which Spitzer photometry
(detections or upper limits) is available.
The remaining objects are addressed in Sect.\ \ref{s_nospitzer}.
The main results are summarised in Table \ref{t_props} and \ref{t_props_2}.
The magnification factors $\mu$ needed to correct for gravitational lensing
are listed in Table \ref{t_mu}.

\subsection{A1835 - \#1}
\begin{figure}[htb]
\centering{\psfig{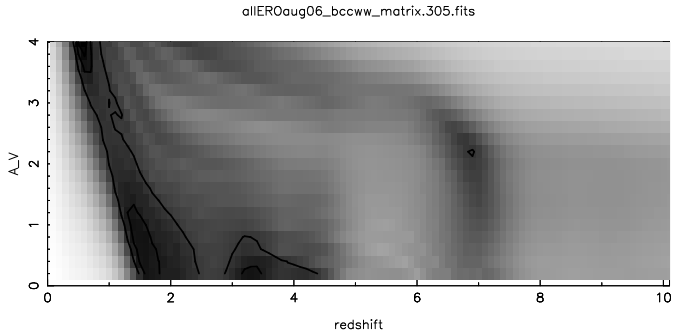}}
\caption{Same as Fig.\ \protect\ref{fig_map_2} for A1835-\#1.}
\label{fig_map_1}
\end{figure}

Overall SED fits for this object are rather degenerate and of poor quality (high \ki2),
showing several minima for its photometric redshift, most of them at
$z \la 3$ (see Fig.\ \ref{fig_map_1}).
Formally, all template groups yield a \ki2\ minimum at $\zfit \sim 0.4$.
%
However, the best fits are of lower quality (higher \ki2) than
for the other objects, since the photometry yields an apparently
somewhat ``non-monotonic'' SED.
Furthermore the photometric redshift is only loosely constrained
as no IRAC photometry is available for channels 3 and 4 (5.8 and 8.0 \micron)
due to blending with other sources.

There is a strong degeneracy in this case between redshift and extinction (Fig.\ \ref{fig_map_1}). 
Two different solutions coexist, one at $\zfit \sim 0.4$ and $A_V \ge$ 3.4, and
another one at $\zfit \sim 1.5$ with $A_V \le 1.2$. The first solution is
strongly degenerate in the age-$A_V$ plane as well, thus providing loose
constraints on the stellar mass: the stellar ages vary quite strongly
from $\sim$ 2.3 Gyr (BC models) to $\sim$ 10 Myr for the Maraston and S04gyr
models, and the corresponding stellar masses range from $\sim 1. \times
10^{10} / \mu \msun$ (BC) to $(0.4-1.1) \times 10^8 / \mu \msun$. 
Even if we consider the solutions around $z \sim 1.5$ there are significant 
uncertainties. E.g.\ the BC and Maraston models require little extinction ($A_V \sim 0.6$),
whereas S04gyr and GRASIL templates indicate a higher extinction (even $A_V \sim 3.8$
for S04gyr!).
Stellar ages of $\sim$ 5.5 Gyr (1 Gyr) are found for the BC and Maraston (S04gyr) models;
the corresponding stellar mass is estimated as $\sim 8.\times 10^{10} \mu \msun$.
For the S04gyr models one obtains $M_\star \sim 6.3 \times 10^{11} \mu \msun$.

\begin{figure}[htb]
\centering\includegraphics[width=8.8cm]{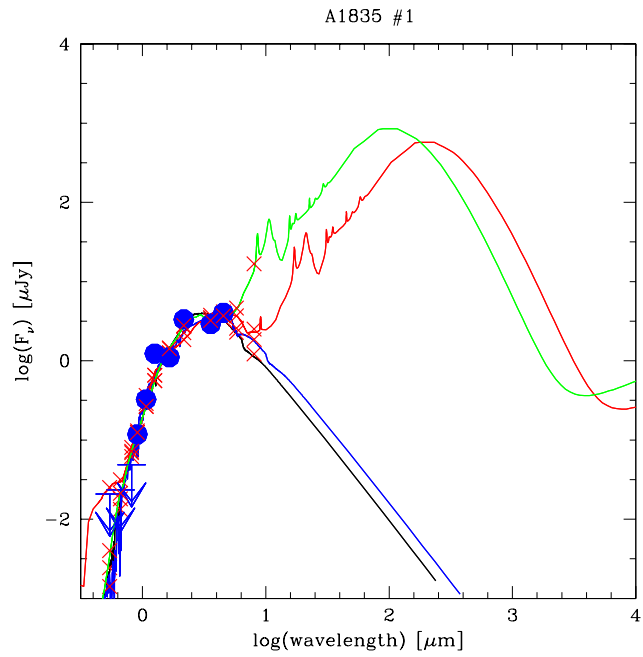}
\caption{A1835-\#1: Comparison between fits with Bruzual and Charlot models
at $z=0.50$ (black) and 1.35 (blue),
and fits with dusty starburst models from GRASIL templates at $z=1.75$ (M82, red) 
and 0.4 (green, M82 plus additional extinction of \av$=3.8$.}
\label{fig_alt_1}
\end{figure}

For illustration we show several SED fits including with the semi-empirical GRASIL 
templates in Fig.\ \ref{fig_alt_1}. The latter allow us in particular to estimate 
the mid-IR to sub-mm flux.
In particular we note that the 24 \micron\ non-detection probably rules out
the very dusty solution at $z \sim 0.4$, rendering $z \sim$ 1.5 more likely.
However, for this object it is clear that the uncertainties
on all derived parameters are large, and larger than for the other objects discussed
here. For this reason the entries in Table \ref{t_props} are left blank for this object.
For comparison we note that the empirical classification based on near-IR colours
would indicate an ``old passive'' object (paper I).

\subsection{A1835 - \#4}

\begin{figure}[htb]
\centering{\psfig{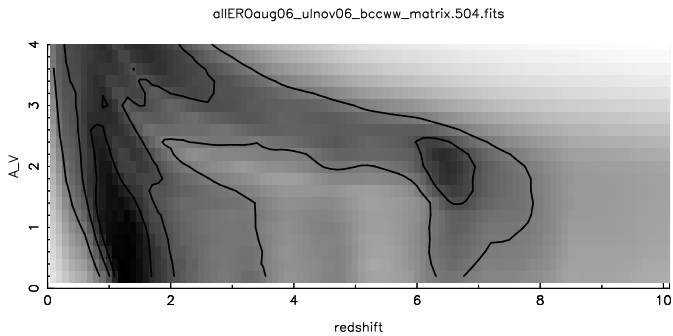}}
\caption{Same as Fig.\ \protect\ref{fig_map_2} for A1835-\#4.}
\label{fig_map_4}
\end{figure}

\begin{figure}[htb]
\centering\includegraphics[width=8.8cm]{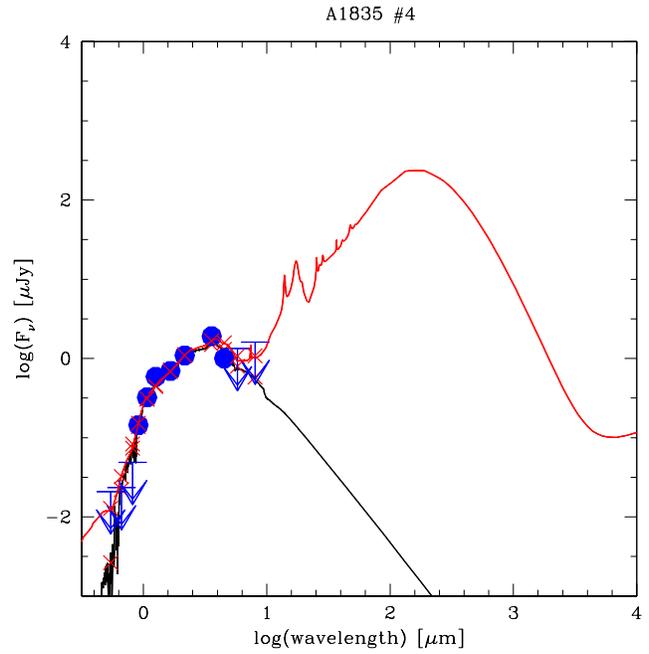}
\caption{A1835-\#4: Comparison between fits with BCCWW galaxy template at $z=1.20$,
and fits with the M82 template from GRASIL at $z=1.24$ (and no additional extinction).
}
\label{fig_alt_4}
\end{figure}

For this object the best fits are consistently found at low redshifts, $\zfit \sim 1.2$
well constrained by the measurement
of the stellar 1.6 \micron\ peak measured in the IRAC channels.
The corresponding \ki2\ maps and the best fit SEDs for this object are 
shown in  Figs.\ \ref{fig_map_4} and \ref{fig_alt_4}.

Best fit templates correspond to bursts of $\sim$ 4.5 Gyr with no extinction 
for Bruzual \& Charlot models or to the elliptical template from CWW,
i.e.\ an old and dust-free galaxy.
With the S04 templates the best fit is of similar quality, yielding
a younger burst age ($\sim 0.6$ Gyr) and some extinction ($A_V \sim 1.6$).
From the BC and S04 model sets the estimated mass is 
$M_\star \sim (1.1-1.7) \times 10^{10} / \mu \msun$, with
the magnification factor $\mu \sim 1.5$ (cf.\ Table \ref{t_mu}). 
If we assume $L_K/M=3.2$ as for SCUBA galaxies (cf.\ Borys \etal\ 2005),
one obtains $M=9.4 \times 10^{9} / \mu  \msun$.
For this object SED fits with Maraston models yield a solution of similar
quality, but a lower redshift of $\zfit \sim 0.8$.
The other fit parameters a burst age of
1.7 Gyr, $A_V=1.2$, and a stellar mass of $4.8 \times 10^9 / \mu  \msun$.

For comparison the best fit to GRASIL templates is found at \zfit\ =1.24 with an M82 template
(and no additional extinction), as shown in Fig.\ \ref{fig_alt_4}.
Although the quality of this fit is less than the ones mentioned above,
we cannot completely rule out the presence of dust. Observations at longer
wavelengths would be needed. Assuming that the M82 template is valid,
we estimate a bolometric luminosity of
to $L_{\rm bol} \sim 2.6 \times 10^{10} / \mu  \lsun$ or a star formation
rate of just $SFR \sim 5 / \mu $ \msunyr.

We note also that the results from the best fit agree with the 
empirical classification as ``old passive'' galaxy based on 
near-IR colours (see paper I).

As a cautionary note, we remind the reader that this object has been
found variable over $\sim$ 1 month in the ISAAC photometry taken in
the $SZ$ band (see Richard \etal\ 2006). The $SZ$ flux adopted here
corresponds to the average between the 2 periods. It is currently
unclear if and to which extent the apparent variability influences the
results derived here.

\subsection{A1835 - \#17}
\begin{figure}[htb]
\centering{\psfig{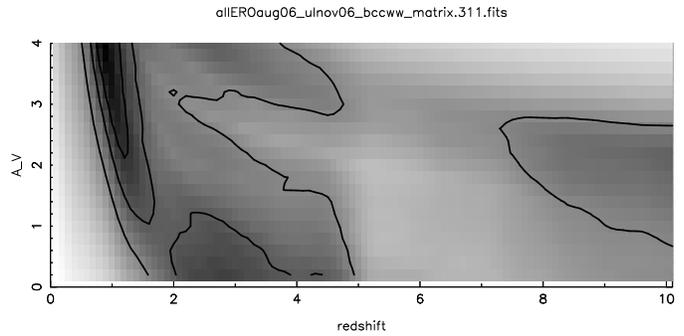}}
\caption{Same as Fig.\ \protect\ref{fig_map_2} for A1835-\#17.}
\label{fig_map_17}
\end{figure}

This object is one of the few for which there is no ambiguity on 
the photometric redshift. 
See \ki2\ map on Fig.\ref{fig_map_17}.
This is mostly due to the fact that the
stellar 1.6 \micron\ peak is clearly observed between the Ks band
and the first two IRAC channels (3.6 and 4.5 \micron).
With nearly all templates one obtains $\zfit \sim 0.8$;
a somewhat lower value of $\zfit \sim 0.69$ is obtained with templates from 
the S04gyr group. 

In all cases a very high extinction ($A_V \sim$ 3--4) is needed.
However, models with very different ages yield fits of similar quality (\ki2):
7.5 Gyr with the BC models, 1 Gyr with Maraston models, and 10 Myr with the 
S04 templates. The SF histories correspond to bursts in all of them.
This age uncertainty is most likely due to the fact that we rely
here on the predictions in the rest-frame spectral range $\sim$ 1.6--2.8 
\micron, where the evolutionary synthesis models are more uncertain
than at shorter wavelengths.
The stellar masses derived from the burst model fits are $\sim (0.6-2.) \times
10^{10} / \mu  \msun$ for the Maraston and BC templates and significantly smaller
from the (younger) S04 template ($M_\star \sim 7. \times 10^7 / \mu  \msun$).
For a value $L_K/M=3.2$ adopted for SCUBA galaxies (cf.\ Borys \etal\ 2005)
one obtains $M_\star=3.1 \times 10^{9} \msun$, much lower than the typical 
masses of SCUBA galaxies.
The magnification factor for this object is $\mu \sim 1.2$ (cf.\ Table \ref{t_mu}).

\begin{figure}[htb]
\centering\includegraphics[width=8.8cm]{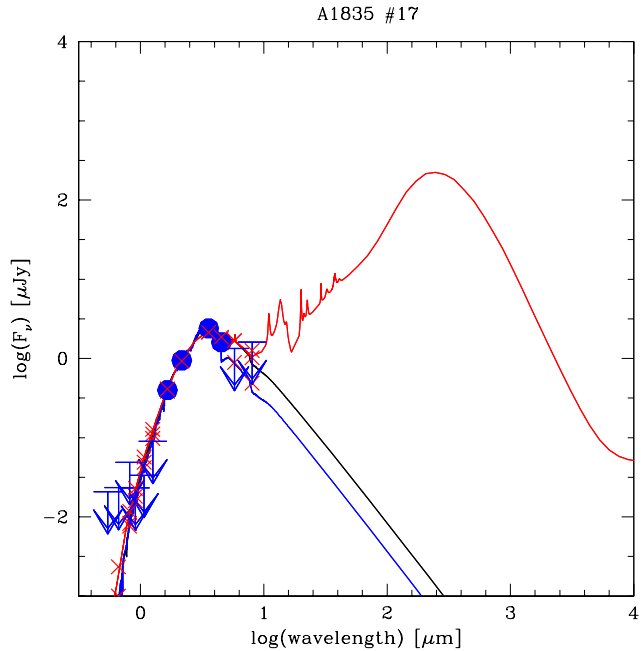}
\caption{A1835-\#17: Comparison between fits with Maraston (2005) templates at $z=0.81$
and \av $=1.$,
and fits with dusty starburst models from GRASIL templates at $z=0.60$ and 0.79
plus additional extinction of \av\ $\sim$ 3.6.
Note that the Maraston template does not include dust emission, which would be
expected for such high an extinction.}
\label{fig_alt_17}
\end{figure}

Using the semi-empirical GRASIL template group the best fits
are found at $\zfit=0.78$ with the SED of the Sbc galaxy M 51 
with an additional optical extinction of \av\ $\sim$ 3.8.
The corresponding mid-IR to sub-mm SED is shown in Fig.\ \ref{fig_alt_17},
with a bolometric luminosity of $5.5 \times 10^9 / \mu  \lsun$ corresponding
to SFR$=0.9 / \mu $ \msunyr.
If this object is indeed a very dusty starburst, which can in principle be verified
with longer  wavelength observations, it is much fainter than SCUBA galaxies
(SMG) at the same redshift --- e.g.\ $\sim$ 4 mag fainter in K (cf.\ Smail \etal\ 2004).

\subsection{AC114-\#1}

\begin{figure}[htbp]
\centerline{\psfig{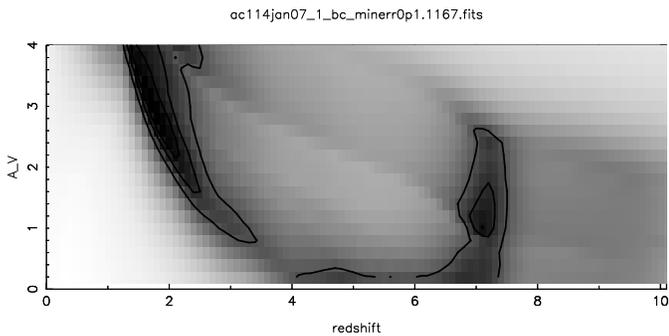}}
\caption{Same as Fig.\ \protect\ref{fig_map_2} for AC114-\#1.}
\label{fig_map_ac1}
\end{figure}

Using the templates from synthesis models the best fits for this object
are found between $\zfit \sim$ 1 and 2.5, with a secondary though less
likely solution at high redshift (see Fig.\ \ref{fig_map_ac1}).
Over the interval $\sim$ 1--2.5 the photometric redshift is actually not 
well determined, since the curvature of the SED measured in the 
4 IRAC bands is small and hence the position of the 1.6 \micron\ peak
-- the main constraint on $z$ -- only loosely constrained.

Both the BC and the Maraston templates give quite similar best fits:
$\zfit=$1.3--1.5, a burst with a maximal age of $\sim$ 3.5--4.5 Gyr, a stellar mass of
$\sim (1.6-2.6) \times 10^{12} / \mu \msun$ and a large extinction
($A_V \sim$ 1.6--2.4). The magnification factor for this object is $\mu \sim 2.2$. 
Best fits with a similar \ki2\ are found using the S04gyr templates
at $\zfit$=1.6 for a burst of 0.9--1.0 Gyr age with $A_V=2.8$
and a stellar mass corresponding to $\sim 1.3 \times 10^{12} / \mu \msun$
\footnote{A lower mass $M_\star=2.4 \times \times 10^{11} / \mu \msun$
is obtained using Eq.\ 1 and $L_K/M=3.2$.}.
The main difference between these models is a lower age and higher extinction
in the latter.

\begin{figure}[htbp]
\centering\includegraphics[width=8.8cm]{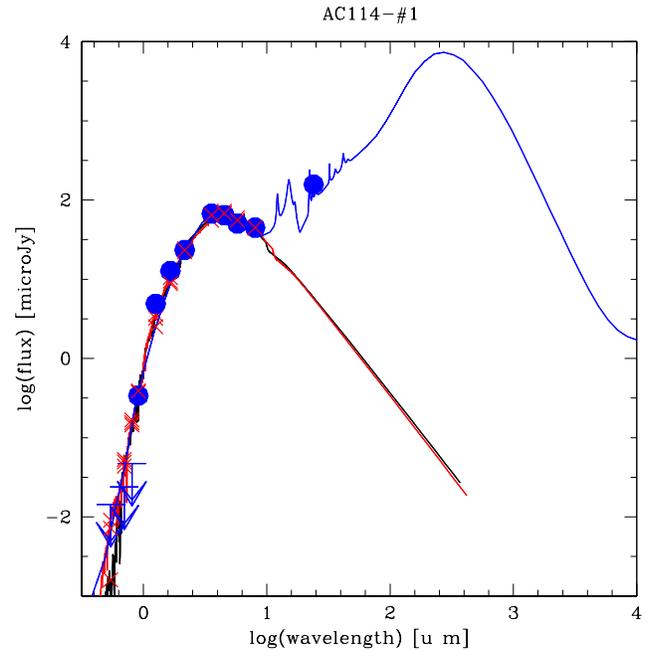}
\caption{AC114-\#1: Best fit SEDs with Bruzual \& Charlot templates
(black line: burst of 4.5 Gyr age and $A_V = 2.4$ at $z=1.3$),
S04gyr templates (red: 1.0 Gyr, $A_V=2.8$, $z=1.6$),
and with GRASIL templates (blue: M52 template + $A_V=3.8$, $z=1.0$).
Note that by construction only the GRASIL templates include dust emission.}
\label{fig_sed_ac114}
\end{figure}

Using the semi-empirical GRASIL templates yields a best fit at $\zfit=$0.9--1.0
for the M51 template. A strong additional extinction of $A_V=3.8$ is required,
and the overall fit is lower quality (higher \ki2) than the fits 
discussed above.
The overall SED resulting from these different fits is shown in Fig.\ 
\ref{fig_sed_ac114}. Interestingly the GRASIL template also reproduces 
quite well the observed MIPS 24 \micron\ flux, although this was not included
in the fit procedure.
In any case the 24 \micron\ flux is a strong indication for 
the presence of dust in this galaxy. In other words solutions with
non negligible extinction at $z \sim$ 0.9--1.5 are favoured by this additional
constraint. 

If at $z=0.97$ the GRASIL template shown in Fig.\ \ref{fig_sed_ac114}
has a bolometric luminosity of $2.8 \times 10^{11} / \mu \lsun$ 
(close to the LIRG range) corresponding to SFR$=48 / \mu \msun$, 
with $\mu \sim 2$ (Table \ref{t_mu}).

From the GRASIL SED shown here we may also expect a fairly strong
sub-mm flux, in the range detectable with current instrumentation.
To the best of our knowledge the southern cluster AC114 has so far not been
observed in this spectral range.

\subsection{Other EROs in A1835 -- galaxies without Spitzer photometry}
\label{s_nospitzer}

The objects treated here are those from Table \ref{t_props_2}
for which contamination by neighbouring sources does not allow us to determine 
photometry from the Spitzer images, i.e.\ the objects \#3, \#10, and \#11 in Abell 1835.
The main properties estimated for \#3 and \#10 are summarised in Table \ref{t_props_2}.

The SED of \#11, shown in Fig.\ \ref{fig_allero},
precludes any reliable photometric redshift estimate; 
above $z \ga 1$ good fits can be found at all redshifts.
For this reason this object is not discussed further.

As for other objects discussed earlier, \#3 and \#10 show a degeneracy
between low and high-$z$, with \ki2\ minima found at $\zfit \sim$ 1--1.5
and 5--6.0 for \#10 (5.--6.5 for \#3). 
However, if at high-$z$ their absolute $Ks$ restframe magnitude is of the order of
-27.3 to -27.6. Such high luminosity objects should be extremely
rare; for this reason we subsequently only consider solutions with
photometric redshifts below 4 and list the properties estimated from the best
fit models.

{\bf A1835-\#3:} 
The best fit is obtained with the BC templates at $\zfit=1.12$ for a burst of 0.5 Gyr with
an extinction of $A_V=0.8$. The corresponding stellar mass is $M_\star \sim 5.1 \times 10^9 / \mu \msun$.
Fits with the Maraston templates yield very similar parameters.
A very similar mass ($ 5.0\times 10^9 / \mu \msun$) is also obtained using Eq.\ 1
and $L_K/M=3.2$ adopted for SCUBA galaxies (Borys \etal\ 2005).

{\bf A1835-\#10:} 
The best fit is obtained with the BC templates at $\zfit=1.23$ for a burst of 0.5 Gyr with
an extinction of $A_V=1.8$. The corresponding stellar mass is $M_\star \sim 9.5 \times 10^9 / \mu \msun$,
basically identical to the one derived following Borys \etal.
Fits with the Maraston templates yield a somewhat younger age (0.25 Gyr), lower extinction
($A_V = 0.4$), and lower mass ($6.4 \times 10^9 / \mu \msun$) at $\zfit=1.14$.

\begin{figure}[htbp]
\centering\includegraphics[width=8.8cm]{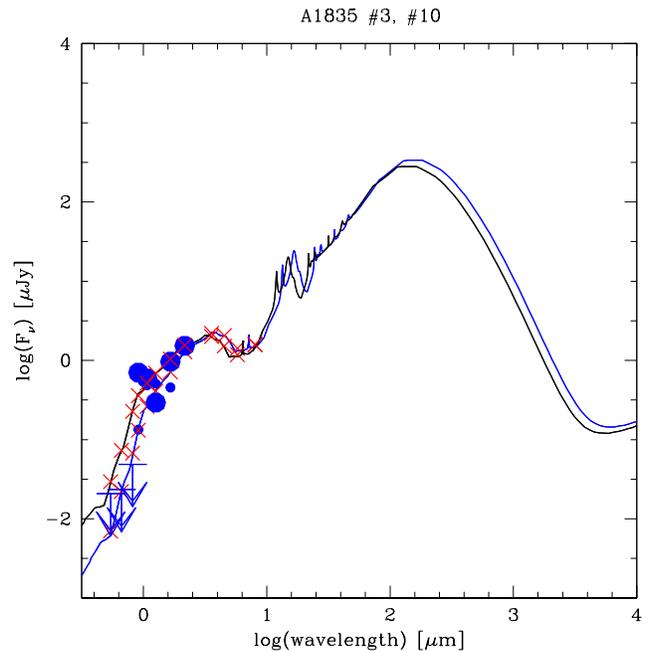}
\caption{Model fits for A1835 \#3 (black) and \#10 (blue) using GRASIL templates
showing predictions for the Herschel/sub-mm/ALMA spectral domain.}
\label{fig_submm_other}
\end{figure}

To illustrate the expected IR to sub-mm SED of these objects we show,
in Fig.\ \ref{fig_submm_other}, the best fits for \#3 and \# 10 obtained with the 
GRASIL templates. 
For reference the best fit redshifts obtained with the GRASIL templates
are $\zfit=0.935$ and 1.165 for \#3 and \#10 respectively, and the SFR 
is $\sim$ 2.4 and 4.3 \msunyr, assuming $\mu=1.4$ for both sources. 
The global SED and IR to sub-mm fluxes of these objects are quite similar to those of \#17.
Remember, however, that the sources \#3 and \#10 are blended with
neighbouring object in the Spitzer images, hence requiring high spatial 
resolution observations to potentially resolve them with future instruments.

\subsection{The importance of IRAC/Spitzer photometry on the photometric redshifts}

\begin{figure}[htbp]
\centering\includegraphics[width=8.8cm]{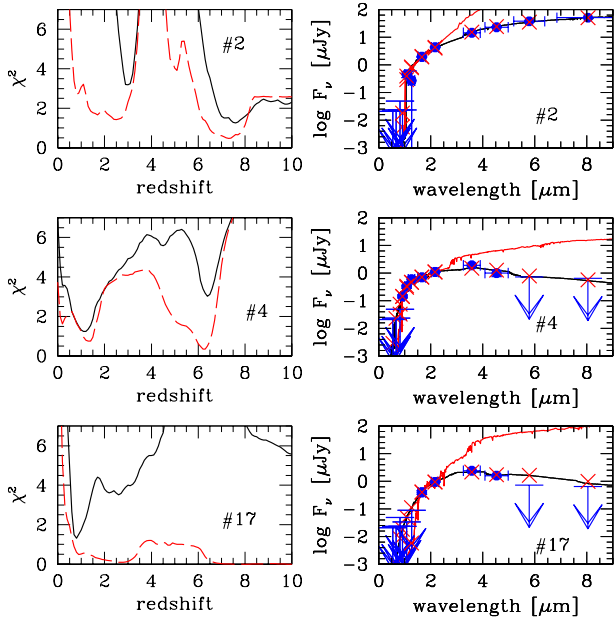}
\caption{Comparison of photometric redshifts derived with or without IRAC photometry for 
selected objects. 
{\bf Left panels:} \ki2\ of the best fitting template versus redshift 
obtained from SED fits including all the photometry (black solid lines), or obtained
from SED fits excluding the IRAC photometric measurements (red dashed lines).
{\bf Right panels:} Best fit SED including all photometry (black solid curve)
and excluding IRAC photometry (red solid). The observations are show by the blue
symbols.
From top to bottom the objects are: A1835-\#2, \#4, and \#17.
These objects illustrate three typical behaviours: 
{\em 1)} degenerate/ambiguous low- and high-$z$ solutions even with IRAC photometry 
(\#2, top panels),
{\em 2)} degenerate/ambiguous low- and high-$z$ solutions whose degeneracy is lifted
thanks to IRAC  photometry (\#4, middle panel), and
{\em 3)} unconstrained photometric $z$ from ground-based photometry, which becomes
well defined low-$z$ solution with IRAC photometry (\#17, bottom panel).}
\label{fig_noirac_pz}
\end{figure}

For the objects with measurable IRAC photometry we have also examined the importance
of this additional information on the SED fits and the resulting photometric redshifts.
Overall this exercise, summarised in Fig.\ \ref{fig_noirac_pz}, shows that the objects 
can be grouped in three ``classes'':

{\bf 1)} Objects showing degenerate/ambiguous low- and high-$z$ solutions even with IRAC photometry 
(cf.\ Fig.\ \ref{fig_alt_2} with A1835-\#2 shown in the top panel, also: A1835-\#1 to some
extend). 
For A1835-\#2, formally the best fit is found
in both cases (i.e.\ with or without IRAC photometry) at $z \sim$ 7.--7.5). However, the
solution at lower $z$ is clearly favoured from arguments on the absolute magnitude of this
object (cf.\ Sect.\ \ref{s_smm}).

{\bf 2)} Degenerate/ambiguous low- and high-$z$ solutions whose degeneracy is lifted
thanks to IRAC photometry (see A1835-\#4, middle panel, 
and also AC114-\#1). In this case the ``curvature'' measured between near-IR and the IRAC
photometry allows one very clearly to locate the 1.6 \micron\ stellar bump due to 
the minimum in the H$^-$ opacity and hence to constrain the galaxy redshift.
A1835-\#1 is intermediate between this case and the following one (3).

{\bf 3)} Unconstrained photometric $z$ from ground-based photometry, which becomes
well defined low-$z$ solution with IRAC photometry (A1835-\#17, bottom panel).
This object, the faintest of our EROs in $H$ and $Ks$, is a $J$-dropout.
Therefore relying only on two ground-based near-IR photometric points results
in a basically unconstrained photometric $z$, as shown in Fig.\ \ref{fig_noirac_pz}
(left bottom panel, dashed line). The two IRAC detections at 3.6 and 4.5 \micron\
together with the upper limits at longer wavelength, are again attributed
to the stellar 1.6 \micron\ bump, making this object a very clear $z \sim 0.81$
galaxy.

In conclusion, if sufficient near-IR ground-based and IRAC/Spitzer photometric
datapoints can be secured and if they reveal a strong enough ``curvature''
such as to constrain the redshifted stellar 1.6 \micron\ bump, SED fitting
of EROs can yield fairly well defined photometric redshifts. 
In other cases, ambiguities between low- and high-$z$ may remain, and other arguments/data
are needed to securely determine their redshift. 
However, given the resulting  bright absolute magnitudes of these objects -- 
if at high-$z$ -- and their large number we consider the ``low-$z$'' (0 $\la z \la$ 3)
solution much more likely for all of the objects studied here. 

Although not included in the SED fitting procedure done with \hyperz, 
a measurement or a stringent upper limit of the 24 \micron\ flux with MIPS
can also provide important constraints, helping e.g.\ to distinguish 
between SED fits with or without strong redenning and therefore indirectly
to rule out certain redshifts (see e.g.\ the cases of A1835-\#2, AC114-\#1).

\section{SED fitting of related objects}
\label{s_other}

\subsection{HUDF-J2 revisited}
Using the same modeling technique and spectral templates described above,
we have examined the optical (\zacs) drop-out object HUDF-J2 discussed by
Mobasher \etal\ (2005) and more recently by Dunlop \etal\ (2006).
This object was also selected by Yan \etal\ (2004) on the basis
of its red 3.6 \micron / \zacs-band flux. Their full sample, named
``IRAC'' EROs (IEROs), will be discussed below (Sect.\ \ref{s_yan}).

\subsubsection{Available photometry}
Here we have used two datasets published for this object.
{\em (i):} JHK photometry from NICMOS and ISAAC, IRAC/Spitzer
data (channel 1--4), and the \zacs\ non-detection from 
Mobasher \etal\ (2005).
In addition the 24 \micron\ flux measured by Mobasher \etal\ (2005)
is also used for comparison, though not included in the \hyperz\
spectral fitting.
{\em (ii):} same data as (i), except for revised values of the optical data
from Dunlop \etal\ (2006) yielding a $B_{435}$ non-detection plus detections 
in $V_{606}$, $i_{775}$ and \zacs.
Where necessary, quoted non-detection limits have been transformed to
$1 \sigma$ limits for a consistent treatment with the \hyperz\
photometric redshift code.

\subsubsection{Results}
\begin{figure}[htb]
\centerline{\psfig{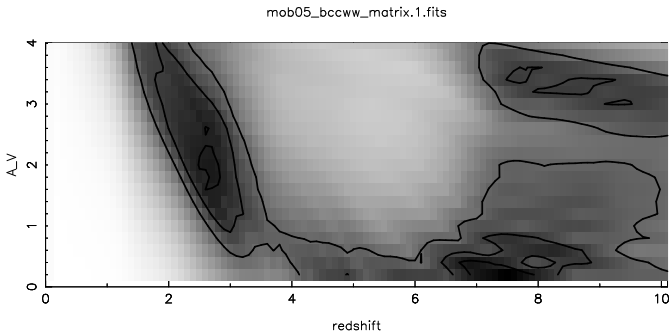}}
\caption{Same as Fig.\ \protect\ref{fig_map_2} for HUDF-J2.
Formally the best fit is found at high redshift (\zfit $\sim$ 6.4--7.4). 
However, for various reasons, including the 24 \micron\ flux 
and the exceptional luminosity of this object if at high $z$, 
we consider it more likely that this galaxy is at $z \sim $ 2.4--2.6.
See discussion in text.}
\label{fig_map_mob}
\end{figure}

Using the photometric dataset (i) from Mobasher \etal\ (2005) we find
quite similar results as these authors, i.e.\ a degeneracy between
solutions at $z \sim 2.5$ plus significant extinction and 
at $z \sim$ 6--7.5 with zero/little extinction, with the formal
best fits lying at high-$z$. This is illustrated by the 
\ki2\ maps shown in Fig.\ \ref{fig_map_mob}.
The main results for this object are summarised in Table \ref{t_mobasher}.

\begin{figure}[htb]
\centering\includegraphics[width=8.8cm]{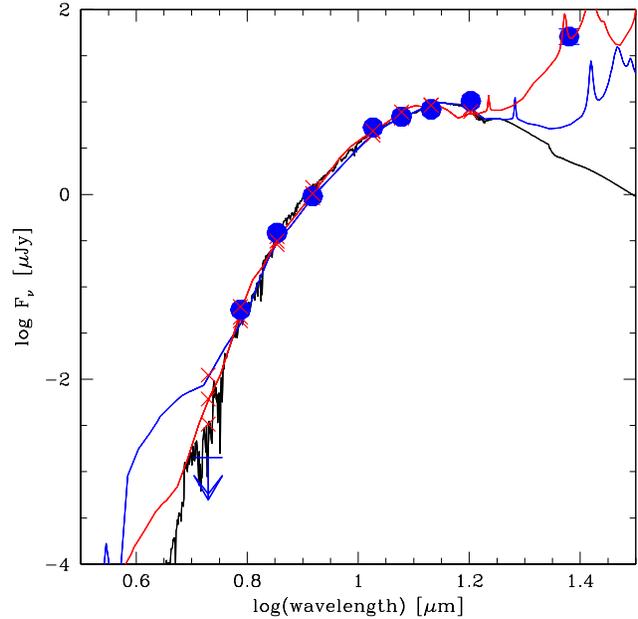}
\caption{$z \sim$ 1.8--2.6 SED fits to the observations of HUDF-J2 from Mobasher \etal\ (2005).
Black line: fit with BCCWW templates at \zfit=2.59.
Red: GRASIL M82 template and additional $A_V=2.8$ at $\zfit \sim 1.8$.
Blue: GRASIL HR10 template without additional extinction at $\zfit \sim 2.3$.
}
\label{fig_sed_mob_nearir}
\end{figure}

Concerning the high-$z$ solutions we note the following.
The best fit \zphot\ depends somewhat on the spectral templates
used; indeed using BCCWW or s04gyr templates we obtain \zfit $\sim$ 7.39 
(\ki2 = 1.5) or \zfit\ $\sim$ 6.5 (\ki2 = 1.2). 
In both cases the best fit is obtained with zero extinction.
With the same templates groups the best fit ``low-$z$'' solutions are 
\zfit=2.59 and $A_V = 1.8$ (\ki2=2.3) for the BCCCWW templates, and
\zfit=2.42 and $A_V = 3.4$ (\ki2=2.6) for the s04gyr templates.
Test calculations have shown, introducing a minimum error
of $\sim$ 0.1--0.15 mag already modifies considerably the \ki2\ map
leading to less well constrained solutions. Given uncertainties in the
determination of measurement errors and uncertainties in matching
photometry from different instruments (mainly NICMOS, ISAAC, and Spitzer)
such errorbars may be more realistic than the small errors quoted 
by Mobasher \etal\ (2005).
A comparison of the photometry of this object published by Yan \etal\
(2004, their object \# 2) and the measurements of Mobasher \etal\ (2005),
showing differences exceeding several $\sigma$ in various bands,
is also illustrative for this purpose.
Given the relatively small differences in the \ki2, the resulting exceptionally 
bright magnitude of this object ( $M^{\rm rest}(Ks) \sim$ --27., cf. Table 
\ref{t_mobasher}), and using the same ``spirit'' as for our EROs, 
we would conservatively favour a ``low-$z$'' interpretation for HUDF-J2
on this basis.
Quantitatively the main weakness of the low-$z$ fits is the slight
excess predicted in the \zacs\ band with respect to the photometry from
Mobasher \etal, as shown in Fig.\ \ref{fig_sed_mob_nearir}.
Actually the flux predicted with the BCCWW template (black line) corresponds to
a $2 \sigma$ detection; a somewhat larger flux is predicted with the
GRASIL templates.

Recently, Dunlop \etal\ (2006) have questioned the very deep \zacs\ non-detection limit
quoted by Mobasher \etal\ (2005), and they have performed manual photometry 
of HUDF-J2 in the optical bands. Their faint detection  in $V_{606}$, $i_{775}$, and \zacs\
alters the balance of the \ki2\ behaviour between low-$z$ and high-$z$ 
favouring solutions at $\zfit = 2.15 \pm 0.3$ (Dunlop \etal\ 2006).
However, imposing a measurement in a prescribed aperture may not be appropriate.
If we use the same constraint and the BC templates we obtain 
$\zfit=2.48$, $A_V=3.0$, and a burst of 0.26 Gyr age, 
in good agreement with Dunlop \etal\ (2006).
In any case, the model fluxes shown in Fig.\ \ref{fig_sed_mob_nearir} in the \zacs\ band 
and at shorter wavelengths, are nicely bracketed by the measurements of
Dunlop \etal\ and Mobasher et al.

\begin{figure}[htb]
\centering\includegraphics[width=8.8cm]{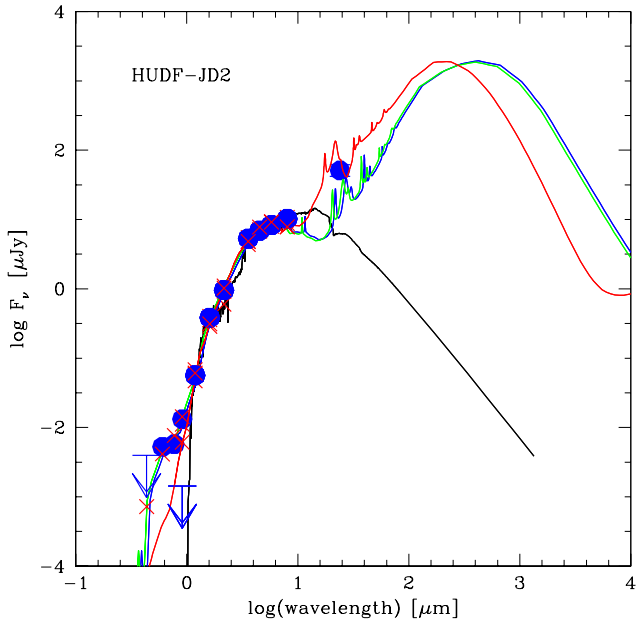}
\caption{SED fits to the observations of HUDF-J2.
The near-IR and IRAC/Spitzer photometry is taken from Mobasher \etal\ (2005);
in the optical we adopt either ({\em (i)} the \zacs\ non-detection from Mobasher \etal\ (2005)
or ({\em (ii)} the revisited photometry from Dunlop \etal\ (2006) yielding 
a $B_{435}$ non-detection plus detections in $V_{606}$, $i_{775}$ and \zacs.
Fitting ({\em (i)} with  BCCWW templates yields a best fit at high-$z$ 
($z \sim 7.4$, black line). 
Fitting ({\em (i)} with GRASIL templates, we obtain a best fit at $z \sim 1.8$
with the M82 template and additional $A_V=2.8$ (red line)
or at $z \sim 2.3$  with the HR10 template (and no additional extinction; shown in blue).
Adopting {\em (ii)} the best fit with GRASIL templates is at $z \sim 2.3$ 
with the HR10 template (and no additional extinction; shown in green).
Note that all ``low-$z$'' solutions using GRASIL templates reproduce 
naturally the observed 24 \micron\ flux.}
\label{fig_sed_mob}
\end{figure}

As qualitatively the SED of HUDF-J2 resembles that of our EROs (in particular
that of A1835-\#2), and given that ``low-$z$'' fits indicate significant
extinction, it is instructive to explore also spectral templates of dusty
objects. Indeed, using the GRASIL templates a reasonable fit (\ki2=4.8) 
to the observations (i)
are found with the M82 template and additional extinction of $A_V =2.8$ at
$\zfit \sim 1.8$ (see red line in Fig.\ \ref{fig_sed_mob}).
Adopting the photometry from (ii), an excellent fit (\ki2=0.9) 
is found with the HR10 template and no additional extinction for
$\zfit = 2.31$ (blue line in Fig.\ \ref{fig_sed_mob}).
Interestingly, both fits reproduce quite well the 24 \micron\ flux
observed by MIPS/Spitzer (Mobasher \etal\ 2005), which was not included
in our fit procedure. 
This lends further credit to the explanation of HUDF-J2 as a 
dust rich galaxy at $z \sim 2.3$.
For comparison, Mobasher \etal\  do not fit the 24 \micron\ emission, and
invoke other components -- possibly an obscured AGN -- to explain this flux.
And this measurement is not considered by Dunlop \etal\ (2006) in their analysis.
As also shown by Fig.\ \ref{fig_sed_mob}, the predicted far-IR to sub-mm flux
of HUDF-J2 should be within reach of existing/future facilities.
Detections in this spectral range should definitely be able to distinguish
between dust-free high-$z$ solutions and the ``low-$z$'' fits favoured here.

From the two GRASIL fits shown in Fig.\ \ref{fig_sed_mob} with the M82 and HR10 templates
at $\zfit=1.83$ and 2.52 respectively we estimate
a bolometric luminosity of $L_{\rm bol} \sim (4.2-6.2) \times 10^{11} \lsun$
(in the LIRG range) corresponding to SFR $\sim$ 72--107 \msunyr.

\subsection{IRAC selected EROs in the HUDF}
\label{s_yan}
To compare our objects with the IRAC selected EROs in the HUDF from Yan \etal\ (2004)
discussed in paper I, we have subjected them to the same quantitative 
analysis using \hyperz. The results from the SED fits are summarised
in Table \ref{t_iero}. Note that for simplicity we list fit results
using Bruzual \& Charlot templates only, although computations
were also carried out using the other template groups. 
Furthermore, to account for possible
mismatches between the photometry from different instruments (ACS, NICMOS,
ISAAC, and IRAC) and to allow for other uncertainties in the error quantification,
we adopt a minimum error of 0.1 mag in all filters.

\subsubsection{Properties of IRAC EROs}
As already noted by Yan \etal, four objects have quite uncertain photometric redshifts.
\#1, 2, and 10 have quite degenerate \ki2\ over a large redshift range.
\#5 yields the worst fit, due to an apparent flux excess at 8 \micron, 
which could be an indication for an AGN contribution.
Therefore we follow Yan \etal\ and exclude these objects from the subsequent
discussion of the average properties of this sample.
However, it is worth mentioning that \# 2 corresponds to the HUDF-J2
object from Mobasher \etal\ (2005) discussed above.

The main difference with the results of Yan \etal\ is that, once allowing
for a reasonable minimum error of say 0.1 mag and including a wide variety
of star formation histories and varying extinction, we are able to obtain
good fits to all objects (except those already mentioned) with standard
Bruzual \& Charlot templates, i.e.\ there is no need for composite stellar
populations as invoked by Yan \etal\ (2004).
In consequence, our best fit ages and extinction differ systematically from
their analysis, as will be discussed below. Otherwise quite similar properties
are derived from our more complete quantitative analysis.

More precisely we find best fit redshifts ranging from 0.6 to 2.8 with a median 
(mean) of 1.6 (1.9). The redshifts difference with Yan \etal\ has
a median value of $\delta z/(1+z) \approx$ -0.07.
In particular we obtain significantly lower redshifts for two objects, \#13 and 14.
However their photometric redshifts  turn out to be quite uncertain
and strongly dependent on the template set used
\footnote{Using the s04gyr (Maraston) templates we find best fits at 
\zfit\ $=$ 2.3 (2.1) and (0.7) 1.3 for \#13 and 14 respectively.}.
We therefore consider them as quite insecure.
Spectroscopic redshifts are available for 3 sources of the Yan \etal\
sample, \#9, \#13, and \#17 (see Daddi \etal\ 2005 and update by Maraston \etal\ 2006).
Except for the uncertain object \#13 just discussed, the agreement with
our photometric redshifts is excellent ($|\delta z| \le 0.05$).

If taken at face value our best fits yield the following average properties
for these IRAC selected EROs
(cf.\ Table \ref{t_iero}):
a median (average) extinction of $A_V =$ 2.0 (2.2),
a median (average) age of 0.5 (1.3) Gyr, and
stellar masses 
$M_\star$ ranging between $10^9$ and $5. \times 10^{11}$ \msun,
with a median (average) mass of $0.5 \times 10^{11}$  ($1.3 \times 10^{11}$) \msun
\footnote{Note that our models assume a Miller-Scalo IMF from 0.1 to 125 \msun,
whereas Yan \etal\ (2004) adopt a Chabrier (2003) IMF.}

Their absolute $M^{\rm rest}_{AB}(Ks)$ covers -20.5 to -24.3 in AB mags,
with a median of -22.3, corresponding to a median of
$M^{\rm rest}_{\rm Vega}(Ks) = -24.2$. This is somewhat brighter than
$M^\star$ from 2MASS (-23.53 cf.\ Kochanek \etal\ 2001)
but similar to $M^\star$ at $z \sim$ 0.6--2.5 from 
Bolzonella \etal\ (2002) and Kashikawa \etal\ (2003)
who find $M^\star_{\rm Vega}(Ks) \sim$ -24.3 to -25.0. 

As already mentioned these average properties are in good agreement with
those derived by Yan \etal\ (2004), except that we find good fits
with templates computed assuming standard star formation histories, i.e.\
that we do not need to invoke other more arbitrary composite stellar populations.
In consequence, the bulk of the fits we obtain correspond to younger stellar ages
and hence to higher extinction than Yan \etal. As already mentioned
by these authors, observations at longer wavelengths (including MIPS imaging)
could help to distinguish between these solutions; in fact, according
to Yan (2006, private communication) 9 of the 17 IERO have been detected 
at 24 \micron\ with fluxes above 20 $\mu$Jy providing support to our
interpretation.

For the subsequent comparison we adopt the properties derived here 
as representative values for the IRAC selected EROs.

\subsubsection{Comparison with our ERO sample}

Empirically our optical drop-out EROs share many properties with
the IRAC selected EROs (cf.\ paper I), in particular similar optical to near-IR and IRAC/Spitzer
colours, and similar \zacs, near-IR and IRAC/Spitzer magnitudes,
after correcting for our median magnification factor corresponding to $\sim$ 0.4 mag.
In contrast to our EROs, which so far are undetected in all bands shortward
of \zacs, the IRAC selected EROs of Yan \etal\ (2004) are detected in optical bands
(BVIz), although at very faint magnitudes (e.g.\ $V \sim$ 27 to 30.).
As the depth of this data, taken from the Hubble Ultra Deep Field, is deeper
than our optical photometry, this is not incompatible with the two samples
being very similar.

The similarity between our EROs and the IRAC selected EROs of Yan \etal\ (2004)
is also evident from the comparison of the derived properties 
(cf.\ Tables \ref{t_props} and \ref{t_iero}), showing similar redshifts, extinction,
stellar ages, and stellar masses.

\begin{table*}[htb]
\caption{Derived/estimated properties for optical dropout ERO galaxies with near-IR and Spitzer 
detections.
Listed are the object ID (col.\ 1), the photometric redshift estimate (col.\ 2), the extinction (col.\ 3),
the type of the best fit template (col.\ 4), the distance modulus corresponding to \zphot\ (col.\ 5),
the absolute $Ks$-band magnitude non-corrected for lensing (col.\ 6), 
the absolute rest-frame $Ks$-band magnitude non-corrected for lensing (col.\ 7),
the estimated stellar mass (from scaling the SED fit or from $M^{\rm rest}(Ks)$ assuming
$L_K/M=3.2$, col.\ 8), the estimated star formation rate  non-corrected for lensing (col.\ 9),
and the age of the stellar population (col.\ 10).
To correct the above mentioned absolute quantities for gravitational magnification the appropriate
magnification factors listed in Table \ref{t_mu} must be used.
}
\begin{tabular}{llllllllllll}
\hline
\hline
Object & \zphot\ & \av\ & template & DM$^a$ & $M^{\rm rest}(Ks)-2.5\log(\mu)$ & Mass$\times \mu$ & SFR$\times \mu$ & stellar age  \\
       &         & [mag]&          & [mag]  & [mag]  & \msun& \msunyr & [Gyr] \\
\hline
A1835-\#1  & $\sim$ 0.4 --1.5    & ?       &    \multicolumn{5}{l}{Fits uncertain -- see text} \\
A1835-\#2  & $\sim$ 2.8--3 & 2.4--3  & young burst & 47.0 & -27.7 & $\sim 1.2 \times 10^{12}$ & $\sim$2100 & $<$ 0.36 \\
A1835-\#4  & $\sim$ 1.2    & 0--1.6  & burst/elliptical & 44.60 & -21.6 & $\sim 1.7 \times 10^{10}$ & $\sim$ 5 & 0.7 to 4.5 \\
A1835-\#17 & $\sim$ 0.7--0.8 & $\sim$ 3.8  & burst     & 43.0 & -21.7 & $\sim 1.3 \times 10^{10}$? & $\sim$ 0.9 &  ? (see text)\\  
\\
AC114-\#1  & $\sim$ 1.3--1.6& $\sim$ 1.6--2.8 & burst & 44.84 & -26.4 & $(1.3-2.6) \times 10^{12}$ &  & $\sim$ 0.9--4.5 Gyr \\
AC114-\#1  & $\sim$ 1.0     & +3.8      & M51         & 44.03 & -25.9 &                      & $\sim$ 48 & \\
\hline
 \multicolumn{5}{l}{$^a$ distance modulus computed for minimum redshift}
\end{tabular}

\label{t_props}
\end{table*}

\begin{table*}[htb]
\caption{Same as Table \ref{t_props} for optical dropout ERO galaxies detected only 
in the near-IR (no Spitzer photometry available). No information is listed for A1835-\#11 due
to its highly uncertain photometric redshift.}
\begin{tabular}{llllllllllll}
\hline
\hline
Object & \zphot\ & \av\ & template & DM & $M^{\rm rest}(Ks)-2.5\log(\mu)$ & Mass$\times \mu$ & SFR$\times \mu$ & stellar age  \\
       &         & [mag]&          & [mag]  & [mag] & \msun& \msunyr & [Gyr] \\\hline
A1835-\#3  & $\sim$ 1.1 & $\sim$ 0.6--0.8 & burst & 44.4 & -22.2 & $\sim 5.1 \times 10^{9}$ & & 0.5 \\
A1835-\#10 & $\sim$ 1.2 & $\sim$ 1.8 & burst & 44.68 & -22.9 & $\sim 9.5 \times 10^{9}$ & & 0.5 \\ 
A1835-\#11 &  ?  \\
\end{tabular}
\label{t_props_2}
\end{table*}

\begin{table*}[htb]
\caption{Same as Table \ref{t_props} for the HUDF-J2 galaxy from Mobasher \etal\ (2005).
To the best of our knowledge no lensing correction has to be applied to this object
($\mu=1$).
}
\begin{tabular}{llllllllllll}
\hline
\hline
Object & \zphot\ & \av\ & template & DM$^a$ & $M^{\rm rest}(Ks)$ & Mass & SFR & stellar age  \\
       &         & [mag]&          & [mag]  & [mag]  & \msun& \msunyr & [Gyr] \\
\hline
HUDF-J2   & $\sim$ 2.4--2.6 & $\sim$ 1.8--3.4& burst & $\sim$46.47 & -24.1 to -24.5 & $\sim 5. \times 10^{11}$& & $\la$ 0.6--2. \\ 
HUDF-J2   & $\sim$ 6.4--7.4 & $\sim$ 0       & burst & 49.03       & -26.9 & $(2.-3.) \times 10^{12}$& & $\la$ 0.7 \\ 
HUDF-J2   & $\sim$ 2.5      &                & HR10  & 46.58       & -23.9 & & 107 & \\
HUDF-J2   & $\sim$ 1.8      & $\sim$ +2.8    & M82   & 45.73       & -23.0 & & 72 \\
\hline
\end{tabular}
\label{t_mobasher}
\end{table*}


\begin{table*}[htb]
\caption{Derived/estimated properties for IRAC selected EROs from Yan \etal\ (2004)
from \hyperz\ fits with Bruzual \& Charlot templates with different star formation
histories, and computed assuming a minimum error of 0.1 mag in all bands.
Note: all magnitudes are in the Vega system ($M(Ks)_{\rm Vega} = M(Ks)_{\rm AB} - 1.871$).
}
\begin{tabular}{llllllllllll}
\hline
\hline
Object & $z_{\rm Yan}$ & \zphot\ & template & age & \av\ & DM & $M^{\rm rest}(Ks)$ & Mass \\
       &               &         &          & [Gyr] & [mag]& [mag]  & [mag]  & \msun& \\\hline
 1* & 3.6 & 2.5 &  burst        & 0.7  & 2.4  & 46.60 &  -25.85    &  3.4e+11\\
 2*a & 3.4 & 2.1 & burst        & 0.7  & 2.8  & 46.12 &  -25.06    &  1.7e+11 \\
 3  & 2.9 & 2.8 &  $\tau=1$ Gyr & 2.6  & 2.0  & 46.82 &   -25.9    & 3.5e+11 \\
 4  & 2.7 & 1.6 &  $\tau=1$ Gyr & 4.5  & 2.0  & 45.43 &  -24.02    & 7.7e+10 \\
 5* & 2.8 & 3.0 &  $\tau=1$ Gyr & 2.3  & 2.2  & 47.04 &  -26.22    & 2.1e+11 \\
 6  & 2.3 & 1.1 &  burst        & 0.01 & 3.8  & 44.40 &   -22.8    &  1.2e+09 \\
 7  & 2.7 & 2.6 &  $\tau=5$ Gyr & 2.6  & 1.8  & 46.69 &  -25.66    & 5.2e+11 \\
 8  & 2.9 & 2.7 &  $\tau=1$ Gyr & 2.3  & 1.4  & 46.71 &  -26.13    & 3.8e+11 \\
 9  & 2.8 & 2.7 &  burst        & 0.4  & 1.0  & 46.76 &  -25.19    & 1.4e+11 \\
10* & 2.1 & 0.9 &  burst        & 0.2  & 3.8  & 43.75 &  -20.58    & 8.3e+08 \\
11  & 2.4 & 1.6 &  burst       & 0.006 & 3.8  & 45.36 &   -23.4    & 6.2e+09 \\
12  & 1.9 & 2.5 &  burst        & 0.4  & 0.6  & 46.55 &  -23.54    & 2.9e+10 \\
13$^b$  & 1.9 & 0.7 &  burst    & 0.2  & 3.8  & 43.00 &  -22.38    & 9.5e+08\\
14$^b$  & 1.6 & 0.6 & $\tau=5$ Gyr & 2.6  & 3.8  & 42.61 & -22.79 & 2.1e+10\\
15  & 2.7 & 2.7 &  $\tau=1$ Gyr & 1.7  & 1.0  & 46.75 &  -24.16    & 5.2e+10 \\
16  & 2.4 & 1.6 &  burst        & 0.01 & 2.4  & 45.44 &  -24.33    & 5.0e+09 \\
17  & 1.6 & 1.7 &  burst        & 0.5  & 0.8  & 45.50 &  -24.38    & 5.1e+10\\
\hline
\multicolumn{9}{l}{$*$ Uncertain photometric redshifts}\\
\multicolumn{9}{l}{ $^a$ same as object HUDF-J2 from Mobasher et al.\ (2005)}\\
\multicolumn{9}{l}{$^b$ \zphot\ redshift likely underestimated (see text)}\\
\end{tabular}
\label{t_iero}
\end{table*}
 
\section{Discussion}
\label{s_discuss}

A fact worth mentioning concerning the objects discussed here
is that one of the sources, AC114-\#1, fulfills the criteria 
commonly used to select $z \sim 6$ galaxies as $i$-dropout.
Indeed for this object $(I_{814}-$\zacs$)_{AB}>2.16$ and it is undetected
at all wavelengths shorter than \zacs. Thus the commonly adopted
$i$ dropout criterion for ACS observations, $(I_{775W}-$\zacs$)_{AB} > 1.5$
(cf.\ Stanway \etal\ 2003, Dickinson \etal\ 2004)
yielding $(I_{775W}-$\zacs$)_{AB} > 1.85$ once transformed to the $I_{814}$ 
filter from WFPC2 (Sirianni \etal\ 2005), is satisfied by this ERO. 
Furthermore, the flux expected in B and V filters from our SED fits is sufficiently 
low ($V_{F606W} \ga 29.2$ in AB mag) that it would remain undetected in 
the GOODS imaging with ACS (cf.\ Giavalisco \etal\ 2004).
Therefore such a low redshift object, with an intrinsic lensing corrected 
magnitude of \zacs$_{AB} \sim 26.$, could potentially contaminate 
$i$-dropout samples. Apparently this is not the case for the sample
of IRAC selected EROs from Yan \etal\ (2004), which if detected 
in \zacs\ are also detected at shorter wavelengths.

Obviously, observations in several filters longward of \zacs\ are 
able to eliminate such objects with very red SEDs out to the IR.
For example, examining $J-H$ Stanway \etal\ (2005) find that their $i_{775W}$-dropout
sample is consistent with unreddened high-$z$ starbursts.
On the other hand, in a recent analysis combining ACS and Spitzer imaging
of the GOODS field, Yan \etal\ (2006) find that $\sim$ 15-21 \% of the 
IRAC detected $i_{775W}$-dropouts have very high flux ratios between 3.6 \micron\ and
the \zacs\ band. These amount to $\sim$ 4 \% of their total $i$ dropout, 
i.e.\ $z \approx 6$  sample.
Our object, AC114-\#1, belongs very likely to the same class of rare objects.

\section{Conclusions}
\label{s_conclude}

We have undertaken a detailed analysis of the stellar populations
and the extinction of a sample of 8 lensed EROs
in Abell 1835 and AC114 from Richard \etal\ (2006) and 
related objects from the Hubble Ultra Deep Field (HUDF).
The analysis includes in particular one known SCUBA galaxy (\smm), 
the $z \sim 6.5$ post-starburst galaxy candidate from Mobasher \etal\ (2005),
and the IRAC selected EROs from Yan \etal\ (2004).

Empirically, these objects share a very red overall SED, similar
colours and magnitudes, and very faint or absent flux in optical bands. 
In particular most of our EROs, originally selected as very red ($R-Ks>5.6$) 
optical drop-out objects by Richard \etal, have been detected with ACS/HST
in deep \zacs\ images, as discussed in Hempel \etal\ (2006, paper I).

The ACS, VLT, and IRAC/Spitzer photometry has been taken from paper I.
To determine photometric redshifts and to simultaneously constrain
the stellar population and extinction properties of these objects,
we have used an updated version of the \hyperz\ code from Bolzonella \etal\ (2000)
including a large number of synthetic, semi-empirical and empirical
spectral templates.

The main results from the SED fitting are the following:
 
\begin{itemize}
\item The SED analysis including near-IR plus IRAC photometry for 5 of our objects,
shows in most cases degenerate solutions between ``low-$z$'' ($z \sim$ 1--3) and
high-$z$ ($z \sim$ 6--7) for their photometric redshifts.
Although formally best fits are often found at high-$z$, their resulting
bright absolute magnitudes, the number density of these objects, and in some
cases Spitzer photometry or longer wavelength observations, 
strongly suggest that all of these objects are at ``low-$z$''.

\item The majority of our lensed objects are best understood by relatively young 
($\la$ 0.5--0.7 Gyr) and dusty starbursts.

\item For 3 of our objects we find indications for strong extinction,
with $A_V \sim$ 2.4--4. 
Among them, the galaxy \smm\ is a known sub-mm emitter and at least one of them,
the ERO AC114-\# 1, is predicted to be in the LIRG category and expected to be 
detectable with current sub-mm instruments.
For the remaining objects, among which 3 show moderate to strong extinction, 
we present predictions of their IR to sub-mm SEDs for future observations 
with APEX, Herschel, and ALMA.

\item  The stellar masses estimated for our objects span a large range from
$\sim 5. \times 10^9 / \mu$ to $1. \times 10^{12} / \mu \msun$,
where $\mu$ is the magnification factor derived from the gravitational lensing
model. Typically one has $\mu \sim 1.2$, with a maximum magnification of $\la 2$.
Where appropriate, star formation rates estimated from the bolometric luminosity
determined from spectral template fitting are SFR $\sim (1-36) / \mu$ \msunyr.
For \smm, the most extreme case, we estimate $\lbol \sim 6. \times 10^{11} \lsun$
and SFR $\sim$ 1000 \msunyr\ for $\zfit \sim 3$ and $\mu=2$.

\item Taking uncertainties and revisions of the photometry of HUDF-J2
as well as accounting for a variety of spectral templates, we
suggest that this object, originally identified as a $z \sim$ 6.5 massive post-starburst,
is more likely a dusty starburst 
at $z \sim$ 2.3--2.6 in agreement with the reanalysis by Dunlop \etal\ (2006).
We show that this explanation also naturally explains the observed 24 \micron\
emission from this object and we predict its IR to sub-mm SED.

\item Using the same methods we have analysed the sample of IRAC selected EROs from 
Yan \etal\ (2004). Both empirically and from our SED fits we find that these objects
show very similar properties to our lensed EROs.
In contrast to Yan \etal\ we do not find strong indications for composite 
stellar populations in these objects. 

\end{itemize}


\begin{acknowledgements}
We've benefited from interesting discussions with numerous colleagues
we'd like to thank here collectively.
We thank Haojing Yan for private communication of MIPS data.
Support from {\em ISSI} (Internation Space Science Institute) in Bern for an 
``International Team'' is kindly acknowledged. 
This work was supported by the Swiss National Science Foundation,
the French {\it Centre National de la Recherche Scientifique},
and the French {\it Programme National de
Cosmologie} (PNC) and {\it Programme National de Galaxies} (PNG).

\end{acknowledgements}


\clearpage

\end{document}